\documentstyle[preprint,aps,eqsecnum, epsfig]{revtex}
\begin{document}
\tighten
\preprint{PITT-97-447; CMU-HEP-97-14; DOE-ER/40682-139;LPTHE-97-}
\draft 
\title{NON-EQUILIBRIUM EVOLUTION OF A `TSUNAMI': DYNAMICAL SYMMETRY
BREAKING} 
\maketitle
\begin{center}
{\bf Daniel Boyanovsky$^{(a)}$, Hector J. de Vega$^{(b)}$, Richard
Holman$^{(c)}$,\\ 
S. Prem Kumar$^{(c)}$, 
Robert D. Pisarski$^{(d)}$} \\ 
{\it (a) Department of Physics and Astronomy, University of Pittsburgh,
PA. 15260,U.S.A}\\
{\it (b) LPTHE, Universit\'e Pierre et Marie Curie (Paris VI) et Denis Diderot
 (Paris 
VII), Tour 16, 1 er. \'etage, 4, Place Jussieu 75252 Paris, Cedex 05, France}
\\  
{\it (c) Department of Physics, Carnegie Mellon University, Pittsburgh, P.A.
15213, U.S.A.}\\  
{\it (d) Department of Physics, Brookhaven National Laboratory, Upton, NY
11973,
 
U.S.A.}  
\end{center}
\begin{abstract}
We propose to study the non-equilibrium features of heavy-ion collisions by 
following the  evolution of an initial state with a large number of quanta with
a distribution around a momentum $ |\vec k_0| $ corresponding to a  
 {\em thin spherical shell} in momentum space, a `tsunami'. An $
O(N) \; ({\vec \Phi}^2)^2 $ model field theory in the large $ N $ limit
is used as a  framework to study the non-perturbative aspects of the
non-equilibrium dynamics including a resummation of the effects of the
medium (the initial particle distribution).  
In a theory where the symmetry is spontaneously broken in the absence of the
medium, when the initial number of particles per correlation volume is chosen
to be larger than a critical  
value the medium effects can restore the symmetry of the initial
state. We show that if one begins with such a symmetry-restored, non-thermal,
initial state, non-perturbative 
effects automatically induce spinodal instabilities leading to a {\em
dynamical} breaking of the symmetry. As a result there is explosive 
particle production and a redistribution of the particles towards
low momentum due to the nonlinearity of the dynamics.  
 The asymptotic behavior displays the onset of Bose condensation of pions and
the equation of state at long times is that of an ultrarelativistic
gas although the momentum distribution is
{\em non-thermal}.  
\end{abstract}

\section{Introduction}

The Relativistic Heavy
Ion Collider (RHIC) at Brookhaven and the Large Hadron Collider (LHC)
at CERN will provide an unprecedented range of energies and luminosities
that will hopefully probe the Quark-Gluon Plasma and Chiral
Phase transitions. The basic picture of the ion-ion collisions in the
energy ranges probed by these accelerators as seen in the center-of-mass frame
(c.m.), is 
that of two highly Lorentz-contracted `pancakes' colliding and leaving
a `hot' region at mid-rapidity with a high multiplicity of 
secondaries\cite{bjorken}.
At RHIC for $Au+Au$ central collisions with typical luminosity of
$10^{26}/cm^2.s$ and c.m. energy $\approx 200 \text{GeV}/n-n$, a multiplicity
of 500-1500 particles per unit rapidity in the central rapidity region
is expected\cite{book1,book2,muller}. At LHC  for head on $Pb+Pb$ collisions
with 
luminosity $10^{27}/cm^2.s$ at c.m. energy $\approx 5 \,\text{TeV}/n-n$, the
multiplicity of charged secondaries will be in the range $2000-8000$
per unit rapidity in the central region\cite{muller}. At RHIC and LHC
typical estimates\cite{bjorken,book1,book2,muller,alam,meyer} of energy
densities 
and temperatures near the central rapidity region are $\varepsilon \approx 1-10
\, \text{GeV}/\text{fm}^3, T_0 \approx 300-900 \text{\,MeV}$.

Since the lattice estimates\cite{muller,meyer} of the transition temperatures
in QCD, both for the QGP and Chiral phase transitions are $T_c \approx
160-200 \text{\,MeV}$, after the collision the central region will be
at a temperature $T>T_c$. In the usual dynamical scenario that one
\cite{bjorken} 
envisages, the  initial state cools off via hydrodynamic expansion through the
phase transition down to a freeze-out temperature, estimated to be $T_F \approx
100 \text{\,MeV}$\cite{satz}, at which the mean free-path of the hadrons is
comparable to the size of the expanding system. 

The initial state after the collision is strongly out of equilibrium and there
are very few quantitative models to study its subsequent evolution. 
There are perturbative and non-perturbative phenomena that contribute
to the processes of thermalization and hadronization. The perturbative
(hard and semihard) aspects are studied via
 parton cascade models which assume that at large energies the nuclei can
be resolved into their partonic constituents and the dynamical evolution
can therefore be tracked by following the parton distribution functions
through the perturbative parton-parton
interactions \cite{wang,geigmuller,geiger,eskola,eskola2}. Parton cascade
models 
(including screening corrections to the QCD parton-parton cross sections)
predict that thermalization occurs on time scales $\approx 0.5\,
\text{fm}/c$\cite{shuryak}. After thermalization, and provided that the
mean-free 
path is much shorter than the typical interparticle separation, further
evolution of the plasma can be described with  boost-invariant relativistic
hydrodynamics \cite{bjorken,cooperfry}. The details of the 
 dynamical evolution {\em between} the parton cascade through hadronization,
and 
eventual description via hydrodynamics is far from clear but will 
require a non-perturbative treatment.
The non-perturbative aspects of particle production and hadronization
typically   
envisage a flux-tube of strong color-electric fields, in  which the field
energy leads to production of $\bar{q}q$ pairs\cite{biro,tube}. Recently the
phenomenon of pair production from strong electric fields in boost-invariant
coordinates was 
studied via non-perturbative methods that address the non-equilibrium
aspects and allow a comparison with hydrodynamics\cite{cooper2}.

The dynamics {\em near} the phase transition is even less understood
and involves physics beyond the realm of perturbative methods. For
instance, considerable
interest 
has been sparked recently by the possibility that disoriented chiral
condensates (DCC's) could form during the evolution of the QCD plasma through
the chiral phase transition \cite{anselm1}-\cite{revs}. Rajagopal and
Wilczek\cite{wilraj} have argued that if the chiral phase transition 
occurs strongly out of equilibrium, spinodal instabilities\cite{boysinglee}
could lead to the formation and relaxation of large pion domains. This
phenomenon could provide a striking experimental signature of the chiral phase
transition and could provide 
an explanation for the Centauro and anti-Centauro (JACEE) cosmic ray
events\cite{lattes}. An experimental program is underway at Fermilab to search
for candidate events\cite{dccexp1,dccexp2}. Most of the 
theoretical studies of the dynamics of the chiral phase transition and
the possibility of formation of DCC's  have focused on initial
states that are in local thermodynamic equilibrium
(LTE)\cite{gavin}-\cite{boydcc}. 

We propose to study the {\em non-equilibrium} aspects
of the dynamical evolution of highly excited initial states by relaxing the
assumption of initial LTE (as would be appropriate for the initial conditions
in a heavy-ion collision). Consider, for example, 
a situation where the relevant quantum field theory is prepared in an initial
state with a particle
distribution
sharply peaked in momentum space around $\vec k_0$ and $-\vec k_0$ where
$\vec k_0$ is a particular momentum. This configuration would be envisaged to
describe two `pancakes' or `walls' of quanta moving in opposite 
directions with momentum $|\vec k_0|$. In the target frame this
field configuration would be seen as a `wall' of quanta moving towards
the target and hence the name `tsunami'\cite{rob}. Such an initial
state is out   
of equilibrium and under time evolution with the proper interacting
Hamiltonian, 
non-linear effects should result in a redistribution of 
particles, as well as particle production and relaxation. The evolution
of this strongly out of equilibrium initial state would be relevant
for understanding phenomena such as formation and relaxation of 
chiral condensates. Starting from such a state and following the complete
evolution of the system thereon, is clearly a formidable problem even within
the framework of an effective field theory such as the linear $\sigma$-model.  

In this article we consider an even more simplistic initial condition, 
where the occupation number of particles
is sharply localized in a thin spherical {\em shell} in momentum space around a
momemtum $ |\vec k_0| $, 
i.e. a {\em spherically symmetric} version of the `tsunami'
configuration. 
The reason for the simplification is purely technical since spherical symmetry 
can be used to reduce the number of equations. Although this is a
simplification of the idealized problem, it will be seen below that the 
features of the dynamics contain the essential ingredients to help us 
gain some understanding of more realistic situations. 

We consider a weakly coupled $\lambda\Phi^4$ theory ($\lambda\sim 10^{-2}$)
with the fields in the vector representation of the $O(N)$ group. 
Anticipating non-perturbative physics, we study the dynamics consistently
in the leading order in the $1/N$ expansion which will allow an analytic 
treatment as well as a numerical analysis of the dynamics. 

 The pion wall
scenario described above is realized by considering an  initial state described
by a Gaussian wave functional with a large number of particles at
$|\vec k_0|$ and a high density is achieved by taking the number of
particles per correlation volume to be very large. As in finite
temperature field theory, a resummation along the lines of the Braaten
and Pisarski \cite{htl} program must be implemented to take into
account the non-perturbative aspects of the physics in the dense
medium. As will be explicitly shown below, the large $N$ limit in the
case under consideration provides a resummation scheme akin to the
hard thermal loop program \cite{htl}.   

The dynamical evolution of this spherically symmetric ``tsunami''
configuration described above reveals many remarkable features: i) In a theory
where the symmetry is spontaneously broken in the absence of a medium, when the
initial state is the $O(N)$ symmetric, high densty, ``tsunami''
configuration, we find that there exists a critical 
density 
of particles depending on the effective (HTL-resummed)  
coupling beyond which spinodal instabilities are induced leading to a
{\em dynamical} symmetry breaking. ii) When these instabilities occur, there is
 profuse production of low-momentum pions (Goldstone bosons) accompanied by a
dramatic 
re-arrangement of the  
particle distribution towards low momenta. This distribution is non-thermal
and its asymptotic behavior signals the onset of Bose condensation of pions.
iii) The final equation of state of
the ``pion gas'' asymptotically at long times is ultrarelativistic despite the
non-equilibrium distributions.

The paper is organized as follows: In Section II we introduce the model under
 consideration and  describe the non-perturbative framework, 
 namely the large $N$ approximation. Section III is devoted to the
construction 
of the wave functional and a detailed description of the initial conditions 
for the problem. The dynamical aspects are covered in Section IV. We first
outline some issues dealing with renormalization and then provide a qualitative
understanding of the time evolution using wave functional arguments. We argue
 that the system could undergo dynamical symmetry breakdown and provide
 analytic estimates for the onset of instabilities.  We present the results of
 our numerical calculations in Section V which confirm the robust features of
 the analytic estimates for a range of parameters. In  
Section VI we analyze the details of symmetry breaking and argue that the long
 time dynamics can be interpreted as the onset of formation of a Bose
 condensate even when the order parameter vanishes.

Finally in Section VII we present our conclusions and future avenues of
study.

\section{The model}
As mentioned in the introduction we consider a  
$\lambda\Phi^4$ theory with $O(N)$ symmetry in the large-$N$ limit with the
Lagrangian,

\begin{equation}
{\cal{L}}=\frac{1}{2}(\partial_\mu\vec{\Phi}).(\partial^\mu\vec{\Phi})
-\frac{m_B^2}{2}(\vec{\Phi}\cdot\vec{\Phi})
-\frac{\lambda}{8N}(\vec{\Phi}\cdot\vec{\Phi})^2
\end{equation}
where $\vec{\Phi}$ is an $O(N)$ vector, $\vec{\Phi}=(\sigma,\vec{\pi})$ and
$\vec{\pi}$ represents the $N-1$ pions,
$\vec{\pi}=(\pi^1,\pi^2,...,\pi^{N-1})$. We then shift $\sigma$ by its
expectation value in the non-equilibrium state

\begin{equation}
\sigma(\vec{x},t)=\sqrt{N}\phi(t)+\chi(\vec{x},t)\;\;\;;
\langle \sigma(\vec{x},t)\rangle = \sqrt{N}\phi(t)\;.
\end{equation}
We refer the
interested reader to 
several  articles which discuss the implementation of the large $N$ limit 
(see for e.g. \cite{largen1,largen2,largen3,frw1,noneq,boydiss,erice97}). 
The $1/N$ series may be generated by introducing an auxiliary field 
$\alpha(x)$ which is an {\em algebraic} function of $\vec{\Phi}^2(x)$, and then
performing the functional integral over $\alpha(x)$ using the saddle point
approximation in the large $N$ limit\cite{largen1,largen2,largen3}. It can be
shown that the leading order terms in the expansion can be  
easily obtained by the following Hartree factorisation of the quantum 
fields\cite{frw1,noneq,boydiss,erice97},

\begin{eqnarray}
&&\chi^4\rightarrow6\langle\chi^2\rangle\chi^2+\text{constant}
\;\;;\;\chi^3\rightarrow3\langle\chi^2\rangle\chi\label{hart1}\nonumber
\\
&&(\vec{\pi}\cdot\vec{\pi})^2\rightarrow 2\langle\vec{\pi}^2
\rangle\vec{\pi}^2-\langle\vec{\pi}^2\rangle^2 +{\cal O}(1/N)
\label{hart3}\nonumber \\ 
&&\vec{\pi}^2\chi^2\rightarrow\vec{\pi}^2\langle\chi^2\rangle 
+\langle\vec{\pi}^2\rangle\chi^2\;\;;\;\vec{\pi}^2\chi\rightarrow
\langle\vec{\pi}^2\rangle\chi\;.\label{hart2}
\end{eqnarray}
All expectation values are to be computed in the non-equilibrium state.
In the leading order large $N$ limit we then obtain,

\begin{eqnarray}
{\cal L} & = &-\frac{1}{2}\vec{\pi}\cdot(\partial^2+{\cal
M}^2_{\pi}(t))\vec{\pi} 
-\frac{1}{2}\chi(\partial^2+{\cal M}^2_{\chi}(t))\chi -\chi V'(\phi(t),t)+ 
\frac{N \lambda}{8}\langle \pi^2 \rangle^2\;,
\label{lagra} \\
{\cal M}^2_{\pi}(t)  & = & m^2_B+ \frac{\lambda}{2} \left[\phi^2(t)+\langle
{\pi}^2 \rangle\right]\;,\label{massoft}\\ 
{\cal M}^2_{\chi}(t) & = & m^2_B+ \frac{\lambda}{2} \left[3\phi^2(t)+\langle
{\pi}^2 \rangle\right]\;,\label{masschioft}\\    
V'(\phi(t),t)& = &\sqrt{N}\left(\ddot{\phi}+
\frac{\lambda}{2}[\phi^2+\langle\pi^2\rangle]\phi+m_B^2\phi\right)\;,
\label{vprime}\\
\langle\pi^2\rangle & = & \langle\vec{\pi}^2\rangle/N\;. \label{expi2}
\end{eqnarray}

This approximation allows us to expand about field configurations that are far
from the perturbative vacuum. In particular it is an excellent tool for
studying the behaviour of matter in extreme conditions such as high 
temperature or high 
density\cite{cooper2,largen1,largen2,largen3,frw1,noneq,boydiss,erice97}.

One way to obtain the non-equilibrium equations of motion is through the
Schwinger-Keldysh Closed Time Path formalism. This is the usual Feynman path
integral defined on a complex time contour which allows the computation of
in-in expectation values as opposed to in-out S-matrix elements. For
details see the references\cite{ctp}.
The Lagrangian density in this formalism is given by

\begin{equation}
{\cal L}_{neq}={\cal L}[\vec{\Phi}^+]-{\cal L}[\vec{\Phi}^-]\;,
\end{equation}
with the fields $\Phi^\pm$ defined along the forward $(+)$ and backward $(-)$
time branches.
The non-equilibrium equations of motion are then obtained by requiring the
expectation 
value of the quantum fluctuations in the non-equilibrium state to vanish
i.e. from the tadpole equations\cite{noneq},

\begin{equation}
\langle\chi^\pm\rangle=\langle\vec{\pi}^{\pm}\rangle=0\;.
\end{equation}
In the leading order approximation of the large $N$ limit, the Lagrangian for
 the $\chi$ field is
quadratic plus linear and the tadpole equation for the $\chi$ leads to the
equation 
of motion for the order parameter or the zero mode

\begin{equation}
\ddot{\phi}+\frac{\lambda}{2}[\phi^2(t)+\langle\pi^2\rangle(t)]\phi(t)+
m_B^2\phi(t)=0\;.
\label{eqnmotion}
\end{equation}
In this leading approximation the non-equilibrium action for the 
$N-1$ pions is,

\begin{equation}
\int d^4x[{\cal{L}}_{\pi^+}-{\cal{L}}_{\pi^-}]=\int d^3xdt\left\{(
-\frac{1}{2}{\vec{\pi}^+} 
\cdot\partial^2{\vec{\pi}^+}-{\cal{M}}_\pi^{+2}(t)\vec{\pi}^+\cdot\vec{\pi}^+)
-(+\longrightarrow-)\right\}\;.
\label{pionlag}
\end{equation}
We have not written the action for the $\chi$ field fluctuations 
because they decouple from the dynamics of the pions in the leading order in
the large 
$N$ limit\cite{frw1,noneq}.

 Having introduced the model and
the non-perturbative approximation scheme the next step is to
construct an appropriate non-equilibrium initial state or density matrix. 

Although one could continue the analysis of the dynamics using the
Schwinger-Keldysh method, we will study the 
dynamics in the Schr\"odinger  representation in terms of wave-functionals
because this will display the nature of 
the quantum states more clearly. We find it convenient
to work with the Fourier-transformed fields defined as,

\begin{equation}
\vec{\pi}(\vec{x},t)=\frac{1}{\sqrt{V}}\sum_{\vec{k}}
e^{-i\vec{k}\cdot\vec{x}}\ 
\vec{\eta}_{\vec{k}}(t)\;,
\end{equation}
where we have chosen to quantize in a box of finite volume $V$ that will be
taken 
to infinity at the end of our calculations. The Hamiltonian
for the pions is  given by

\begin{equation}
H_{\pi}=\sum_{\vec{k}}\left(\frac{1}{2}\vec{\Pi}_{\vec{k}}\cdot
\vec{\Pi}_{-\vec{k}}+\frac{1}{2}\omega_k^2(t)\vec{\eta}_{\vec{k}}
\cdot\vec{\eta}_{-\vec{k}}\right)-\frac{N\lambda}{8} \left(\sum_{\vec k}
\langle\vec{\eta}_{\vec{k}} \cdot\vec{\eta}_{-\vec{k}} \rangle\right)^2\;,
\label{hamilt} 
\end{equation} 
where 
\begin{equation}
\omega_k^2(t)=\vec{k}^2+{\cal{M}}^2_{\pi}(t)
\label{omegadef}
\end{equation}
is the effective time dependent frequency  and ${\cal{M}}^2_{\pi}(t)$ is given
by Eq.(\ref{massoft}). To leading order in the large 
$N$ limit the theory becomes Gaussian and the non-linearities are encoded
in a self-consistency condition, since the frequency (\ref{omegadef})
depends on $\langle \vec{\pi}^2 \rangle$ and this expectation value
is in the time dependent state, as displayed by the set of equations
(\ref{massoft}-\ref{expi2}).

\section{the initial state}
As stated in the introduction, our ultimate goal is to model and study
the non-equilibrium aspects of the evolution of an initial, highly
excited  state that  relaxes  following high energy,
heavy-ion collisions. An idealized description of the associated physics  would
be to consider two wave packets made up of very high
energy components representing the heavy ions  and moving with a highly
relativistic momentum toward each other. The goal would be to follow  
the dynamical evolution of the wavefunctionals corresponding to this
situation, thus clearly elucidating the non-equilibrium features involved 
in the phase transition processes following the interactions of the wave
packets. This initial state could be described by a
distribution of particles, sharply peaked around some special values
$\vec{k}_0$ and $-\vec{k}_0$ in momentum space. The evolution of this state
then 
follows from the functional Schr\"odinger equation. 

 Even with the simplification
of a scalar field theory such a program is very ambitious and beyond
the present numerical capabilities. One of the major difficulties is that
selecting one particular momentum breaks rotational invariance and the
evolution equations depend on the direction of wave vectors even in
the Gaussian approximation. (This statement will become clear below).

In this article however, we choose to study a much simpler  description of the
initial state which is characterized by a high 
density  particle distribution  in a thin spherical `shell' in momentum space
. We propose an initial particle distribution  that has 
 support concentrated at $|\vec{k}_0|$. 
This particular state does not provide the necessary geometry for a 
heavy ion collision, however it does describe a situation in which
initially there is a large multiplicity of particles in a small momentum
`shell', there is no special beam-axis and the pions
are distributed equally in all directions {\em with a sharp spatial
momentum}. This is a rotation invariant state that describes a highly 
out of equilibrium situation and that will relax during its time evolution (a 
spherical ``tsunami''). 

\subsection{The Wave Functional:} 
Since in the leading order approximation in the large $N$ expansion the theory
has become Gaussian (at the expense of a self-consistency condition),
we choose a  Gaussian ansatz for the wave-functional at $t=0$.
The reason for this choice is that upon time evolution this wave functional
will remain Gaussian and will be identified with a
squeezed state functional of pions. 
\begin{equation}
\Psi(t=0)=\Pi_{\vec{k}}{\cal{N}}_{k}(0)\exp
\left[-\frac{A_{k}(0)}{2}\;
\vec{\eta}_{\vec{k}}\cdot\vec{\eta}_{-\vec{k}}\right]\;.\label{wavefunc0}
\end{equation}
This state  will then evolve according to the
Hamiltonian given in Eq. (\ref{hamilt}) which is essentially a harmonic
oscillator Hamiltonian with self-consistent, time-dependent frequencies. The
functional Schr\"odinger equation is given by 

\begin{equation}
i\frac{\partial\Psi}{\partial t}=H\Psi. \label{timedep}
\end{equation} 
The last term in the Hamiltonian (\ref{hamilt}) which is independent of the
fields (a time dependent `vacuum energy term') can be absorbed in 
an overall time dependent phase of the wave functional. Removing this 
term by a phase redefinition, the functional Schr\"odinger equation 
becomes
\begin{equation}
i\dot{\Psi}[\eta]=\sum_{\vec{k}}\left[
-\frac{\hbar^2}{2}\frac{\delta^2}
{\delta\vec{\eta}_{\vec{k}}\delta\vec{\eta}_{-\vec{k}}}
+\omega_{k}^2(t)\; \vec{\eta}_{\vec{k}}\cdot\vec{\eta}_{-\vec{k}}\right]
\Psi[\eta]\label{schrod}
\end{equation}
which then leads to a set of differential equations for the covariance $A_{k}$.
The time dependence of the normalization factors 
${\cal N}_k$ is completely determined by that of the $A_{k}$ as
a consequence of unitary time evolution. The state for arbitrary time
$ t $ takes then the form:
\begin{equation}
\Psi(t)=\Pi_{\vec{k}}{\cal{N}}_{k}(t)\exp
\left[-\frac{A_{k}(t)}{2}\;
\vec{\eta}_{\vec{k}}\cdot\vec{\eta}_{-\vec{k}}\right]\; .
\label{wavefunct}
\end{equation}

The evolution equations for
the covariance are obtained by taking the functional derivatives and
comparing powers of $\eta_{\vec k}$ on both sides. We obtain the following
evolution equations\cite{boydcc,frw1}

\begin{eqnarray}
i\dot{A}_{k}(t) & = & A_{k}^2(t)-\omega_{k}^2(t),
\label{riccati}\\
{\cal N}_k(t) & = & {\cal N}_k(0) \exp\left[\int^t_0 A_{Ik}(t')dt'\right]\;,
\label{norma}
\end{eqnarray}
with $A_{k}= A_{Rk}(t)+iA_{Ik}(t)$. The equal time two-point correlation
function in the time evolved non-equilibrium state is given by 

\begin{eqnarray}
\langle\vec{\eta}_{\vec{k}}\cdot\vec{\eta}_{-\vec{k}}\rangle
&&=\frac{<\Psi|\; \vec{\eta}_{\vec{k}}\cdot\vec{\eta}_{-\vec{k}}\; |\Psi>}
{<\Psi|\Psi>}\nonumber\\
&&=\frac{\int[{\cal D}\vec{\eta}_{\vec{q}}]\; (\vec{\eta}_{\vec{k}}
\cdot\vec{\eta}_{-\vec{k}})\; \Pi{\cal N}_{\vec{q}}\; 
\exp\left[-\frac{A_q(t)}{2}
\; \vec{\eta}_q\cdot\vec{\eta}_{-q}\right]}{\int[{\cal D}
\vec{\eta}_{\vec{q}}]\;  \Pi{\cal N}_{\vec q}\; \exp\left[-\frac{A_q(t)}{2} 
\vec{\eta}_q\cdot\vec{\eta}_{-q}\right]}\nonumber \\
&&=\frac{N}{2A_{Rk}(t)}\;,
\end{eqnarray}
leading to the self-consistency condition
\begin{equation}
\langle \pi^2 \rangle(t) = \sum_k \frac{1}{2 A_{Rk}(t)}\;. \label{selfish}
\end{equation}

Formally, one can also represent these two-point equal time correlators in
terms of  functional integrals over the closed time path contour where the
initial state is chosen to be the Gaussian functional described above. However
the explicit 
and rather simple ansatz for the wave functional enables one to obtain the
two-point functions directly in a rather straightforward manner. Moreover, the
wave functional approach will permit a much clearer understanding of the
physics of the problem. The Ricatti equation (\ref{riccati}) can be cast in a
simpler form by writing $A_k$ in terms of new variables $\phi^*_k$ as 

\begin{equation}
A_k(t)=-i\frac{\dot{\phi}^*_k(t)}{\phi^*_k(t)}\;,\label{phidef}
\end{equation}
leading to the simple equation for the new variables
\begin{equation}
\ddot{\phi}^*_k+\omega_k^2(t)\;\phi^*_k=0\;.\label{phidiff}
\end{equation}
In terms of these mode functions we find that the real and imaginary parts of
the covariance $A_k$ are
given by  

\begin{eqnarray}
A_{Rk}(t)& = & \frac{i}{2}
\frac{\dot{\phi}_k\phi^*_k-
\dot{\phi}^*_k\phi_k}{|\phi_k|^2},\label{areal}
\\
A_{Ik}(t) & = & - \frac{d}{d t}\ln|\phi_k(t)|^2 \; . \label{aima}
\end{eqnarray}

From the differential equation for the $\phi_k(t)$  given by
Eq. (\ref{phidiff}) it is clear that the combination that appears in
the numerator of Eq. (\ref{areal}) is 
the Wronskian $\Omega_k$ of the differential equations and will consequently be
determined  
from the initial conditions alone. The expression for the quantum fluctuations
$\langle\pi^2\rangle=\langle\vec{\pi}^2\rangle/N$ is given by, 

\begin{equation}
\langle\pi^2\rangle(t)=\int \frac{d^3k}{(2\pi)^3}\; 
\langle\eta_{\vec{k}}(t)\eta_{-\vec{k}}(t)\rangle
=\int \frac{d^3k}{(2\pi)^3} \frac{|\phi_k(t)|^2}{2\Omega_k}\; .
\label{fluct}
\end{equation}

The mode functions $\phi_k$ have a very simple interpretation: they
obey the Heisenberg equations of motion for the
pion fields obtained from the Hamiltonian (\ref{hamilt}). Therefore
we can write the Heisenberg field operators as
\begin{equation}
\vec{\pi}(\vec x,t) = \frac{1}{\sqrt{V}}\sum_k \frac{1}{\sqrt{2\Omega_k}}\left[
\vec{a}_k\; \phi_k(t) \;
e^{i\vec k \cdot \vec x}+\vec{a}^{\dagger}_k \;\phi^*_k(t)\;
e^{-i\vec k \cdot \vec x} \right]
\end{equation}
where $\vec{a}_k \; , \; \vec{a}^{\dagger}_k$ are the time independent
annihilation and creation operators with the usual Bose commutation relations. 

\subsection{Initial Conditions:}

Within this Gaussian ansatz for the wave functional, the initial
conditions are completely determined by the initial conditions on the
mode functions $ \phi_k(t) $. In order to physically motivate the initial
conditions we now establish the relation between the particle number
distribution and these mode functions. 

Since, in a time dependent situation there is an ambiguity in the definition
of the particle number, we {\em define} the particle number with respect to 
the eigenstates of the instantaneous Hamiltonian (\ref{hamilt}) at
the {\em initial time}, i.e.

\begin{eqnarray}
\hat{n}_k=&&\frac{1}{ \omega_k(0)}\left(-\frac{1}{2}
\frac{\delta^2}{\delta\vec{\eta}_{\vec{k}}\delta\vec{\eta}_{-\vec{k}}}+
\frac{\omega_k^2(0)}{2}\vec{\eta}_{\vec{k}}\cdot\vec{\eta}_{-\vec{k}}\right)-
\frac{1}{2}\nonumber \\
=&&\frac{1}{ \omega_k(0)}\left[\frac{1}{2}\vec{\Pi}_{\vec{k}}\cdot
\vec{\Pi}_{-\vec{k}}+\frac{\omega_k^2(0)}{2}
\vec{\eta}_{\vec{k}}\;\cdot\vec{\eta}_{-\vec{k}}\right]-\frac{1}{2}\;.
\label{numberofparts} 
\end{eqnarray}
Here, $\omega_k(0)$ is the  frequency (\ref{omegadef}) evaluated at $t=0$,
i.e. the curvature of the potential term in the functional 
Schr\"odinger equation (\ref{schrod}) at $t=0$ and provides a
definition of the 
particle number (assuming that $\omega_k^2(0)>0$). The expectation value of the
number operator in the time evolved state is then 

\begin{eqnarray}
n_k(t)=&&<\Psi|\hat{n}_k|\Psi>
=\frac{[A_{Rk}(t)-\omega_k(0)]^2+A_{Ik}^2(t)}{4\,\omega_k(0)\,A_{Rk}(t)}
\label{expvaln}\\ 
=&&\frac{\Delta_k(t)^2+\delta_k(t)^2}{4[1+\Delta_k(t)]}\; ,\label{numbexp1}
\end{eqnarray}
where $\Delta_k$ and $\delta_k$ are defined through the relations,
\begin{equation}
A_{Rk}(t)=\omega_k(0)\;[1+\Delta_k(t)]\;\;\;;A_{Ik}(t)=\omega_k(0)
\;\delta_k(t)\;. \label{deltadef}
\end{equation}
In terms of the mode functions $\phi_k$ and $\dot{\phi}_k$ the expectation
value of the number operator is given by 
\begin{equation}
n_k(t)=\frac{1}{4\;\Omega_k\;{\omega_k(0)}}
\left[|\dot{\phi}_k(t)|^2
+\omega^2_k(0)|\phi_k(t)|^2\right]-\frac{1}{2}\;.\label{numbexp2}
\end{equation}
The quantity $\delta_k(t)$ appears as the phase of the wave function and will
be chosen to be zero at $t=0$ for simplicity,

\begin{equation}
A_{Ik}(0)=0\;\;\;;\quad\delta_k(0)=0.
\end{equation}
Assuming $\delta_k(0)=0$, the initial conditions on the $\phi_k(0)$
variables can 
be obtained at once from  Eq. (\ref{phidef}) and are found to be, 
\begin{equation}
\dot{\phi}^*_k(0)=i\Omega_k=i\omega_k(0)\;[1+\Delta_k(0)]\;\;
\;;\phi^*_k(0)=1\;,
\end{equation}
where $\Omega_k$ is the Wronskian $ W[\phi_k(t)^*,\phi_k(t)] $.
Hence the wave functional of the system at $t=0$ can be specified completely
(up to a phase) by the single function $\Delta_k(0)$. 

Using Eq. (\ref{numbexp1}) one can easily solve for $\Delta_k(0)$ in terms of
the initial particle spectrum  
\begin{equation}
\Delta_k\equiv\Delta_k(0)=2[n_k(0)\pm\sqrt{n_k(0)^2+n_k(0)}]\;.\label{Deltak}
\end{equation}
Which of the two solutions will give us interesting physics is a more subtle
question that we shall address in the next section when we discuss the dynamics
of the problem.  

Before moving on to the description of the dynamics let us briefly
summarize what we have done. We proposed a rather simple description of
a large multiplicity, high energy particle collision process by preparing  an
initial state with an extremely high number density of particles concentrated
at momenta given by $|\vec{k}|=k_0$. Consistent with the 
leading order in a large $N$ approximation, we chose a Gaussian ansatz for
our wave functional, parametrized by the variables $\phi_k(t)$,
$\dot{\phi}_k(t)$  (or alternatively $\Delta_k(t)$ and $\delta_k(t)$)
and the initial conditions on 
these variables are determined completely by the choice of the particle
distribution function $n_k(0)$ at $t=0$ (Eq. (\ref{Deltak})). The next step
will be to obtain the {\em renormalized} equations of motion and then to study
the dynamics analytically as well as numerically. 

\section{The Dynamics}

 From the discussion in the previous sections we see that the following set of
equations for the order parameter $\phi(t)$ and the mode functions $\phi_k(t)$
must be solved self-consistently in order to study the dynamics: 

\begin{eqnarray}
&&\frac{d^2\phi(t)}{dt^2}+\left(m_B^2+\frac{\lambda}{2}[\phi^2(t)+
\langle\pi^2\rangle_B]\right) 
\phi(t)=0\; ,\label{zeromode}\\
&&\frac{d^2\phi^*_k(t)}{dt^2}+\left(k^2+m_B^2+\frac{\lambda}{2}[\phi^2(t)+
\langle\pi^2\rangle_B]\right)\phi^*_k(t)=0\;,\label{kmodes}\\
&&\phi^*_k(0)=1\;\;\;;\dot{\phi}^*_k(0)=i\Omega_k\;, \label{inicon}
\end{eqnarray}
with the self-consistent condition
\begin{equation}
\langle\pi^2\rangle(t)_B=\int \frac{d^3k}{(2\pi)^3}
\frac{|\phi_k(t)|^2}{2\Omega_k}\; .\label{selfcons}
\end{equation}
The quantities in the above equations must be renormalized. This is achieved by
first demanding that all the equations of motion be finite and then absorbing
the  divergent pieces into a redefinition of the mass and coupling constant
respectively, 

\begin{equation}
m_B^2+\frac{\lambda}{2}[<\pi^2>(t)_B+\phi^2(t)]
=m_R^2+\frac{\lambda_R}{2}[<\pi^2>(t)_R+\phi^2(t)] = {\cal M}^2_{R\;\pi}(t)\;.
\label{renormass}
\end{equation}
A detailed derivation of the renormalization prescriptions requires a WKB
analysis of the mode functions $\phi_k(t)$ that reveals their ultraviolet
properties. Such an analysis has been performed elsewhere \cite{frw1,noneq}. In
summary 
the mass term will absorb quadratic and logarithmic divergences while the
coupling constant will acquire a logarithmically divergent
renormalization\cite{frw1,largen1}. In particular        
\begin{equation}
<\pi^2>_R(t) = \int \frac{d^3k}{(2\pi)^3}\left\{
\frac{|\phi_k(t)|^2}{2\Omega_k}-\frac{1}{2k}+\frac{\theta(k-\kappa)}{4k^3}
{\cal M}^2_{R\;\pi}(t) \right\}
\label{renofluc}
\end{equation}
with $\kappa$ an arbitrary renormalization scale.  

Introducing the effective mass of the particles at
the initial time as

\begin{equation}
M^2_R = {\cal M}^2_{R\; \pi}(t=0)\; , \label{inimass}
\end{equation}
we recognize that this effective mass has contributions from the
non-equilibrium particle distribution and is the analog of the hard-thermal
loop (HTL) resummed effective mass in a scalar field theory. Recall, however,
that the initial distribution is {\em not} thermal. 
In a scalar theory, the HTL effective mass is obtained by summing the daisy and
superdaisy diagrams \cite{parwani} which 
is precisely the resummation implied in the leading order in the large $N$
approximation. To 
see this more clearly consider the case in which the order parameter
vanishes, i.e. $\phi(t) \equiv 0$; then the effective mass at the
initial time can be written as a gap equation

\begin{eqnarray}  
M^2_R=&&m_B^2+\frac{\lambda}{4\pi^2}
\int \frac{k^2dk}{2\Omega_k(0)}\label{dressedmass}\\\nonumber\\\nonumber
=&&m_B^2+\frac{\lambda}{4\pi^2}
\int \frac{k^2dk}{2\omega_k(0)}+
\frac{\lambda}{4\pi^2}\int
\frac{k^2dk}{2\omega_k(0)}\left[-\frac{\Delta_k(0)}{1+\Delta_k(0)}\right]
\end{eqnarray}
where we have used the relation $ \Omega_k(0)=\omega_k(0)(1+\Delta_k) $.
The second term in the above expression is the usual contribution obtained
at zero temperature (and zero density) for the (self-consistent) renormalized
mass parameter $M_R^2$ i.e the 
one loop tadpole (with $\omega_k(0)=
\sqrt{k^2+M^2_R}$). The third term contains the non-equilibrium effects
associated 
with the particle distributions and vanishes when $n_k \rightarrow 0$. This
term is finite (since $n_k(0)$ is assumed to be localized within a small range
of momenta).  For a given distribution $n_k(0)$, the solution to the
self-consistent gap  
equation (\ref{dressedmass}) gives the effective mass, dressed by the 
medium effects. This is indeed very similar to the finite temperature case in
which the tadpole term provides a contribution $\propto \lambda T^2$ in the
high temperature limit. We will see later that a term very similar to this can
be extracted in the limit in which the distribution $n_k(0)$ is very large.

Since the relevant scale is the quasiparticle mass $M_R$,
we will choose to take $M^2_R > 0$ to describe an initial situation in
which the $O(N)$ symmetry is unbroken. 

It is convenient
for numerical purposes to introduce the following dimensionless quantities: 

\begin{eqnarray}
&&q=\frac{k}{M_R} ;\;\;\;\;{\tau}=M_R\;t ;\;\;\;\; \varphi^2(\tau)=
\frac{\lambda_R\; \phi^2(t)}{2M^2_R}\; ; \; g=\frac{\lambda_R}{8\pi^2}\; ; \; 
W_q= \frac{\Omega_k}{M_R}
\nonumber\\
&& g\Sigma(\tau) = \frac{\lambda}{2M^2_R}\left[\langle \pi^2 \rangle_R(t)
-\langle\pi^2\rangle_R(0)\right]\;. \label{dimless} 
\end{eqnarray}
In terms of these dimensionless quantities the equations of motion
(\ref{zeromode}, \ref{kmodes}) with 
the initial conditions (\ref{inicon}) become

\begin{eqnarray} 
&&\left[\frac{d^2}{d\tau^2}+1+\varphi^2(\tau)-\varphi^2(0)+
g\Sigma(\tau)\right]  
\varphi(\tau)=0\;,\label{zeromodeeq}
\\
&&\left[\frac{d^2}{d\tau^2}+q^2+1+\varphi^2(\tau)-\varphi^2(0)+
g\Sigma(\tau)\right]\phi_q(\tau)=0 \; \; ; \; \; \phi_q(0)=1 \;; \;
\dot{\phi}_q(0)= -iW_q\;,  
\label{modeeq} \\
&&g\Sigma(\tau) =  g \int q^2dq \left\{
\frac{|\phi_q(\tau)|^2-1}{ W_q}+\frac{\theta(q-1)}{2q^3}
\left({{{\cal M}^2_{R\;\pi}(\tau)}\over {M^2_R}}-1\right) \right\}\;,
\label{gsigma} 
\end{eqnarray} 
where we have chosen the renormalization scale $\kappa = M_R$ for simplicity. 
 
In order to make our statements precise in the analysis of the
spherically symmetric ``tsunami'', we will assume   the initial particle 
distribution to be Gaussian and peaked at some value $q_0$ and width 
$\xi$ so that 

\begin{equation}
n_q(0)=\frac{N_0}{I}\exp\left[-\left[\frac{q-q_0}{\xi}\right]^2\right],
\label{inidistbn} 
\end{equation}
where  $N_0$ is the total number
of particles in a  correlation volume $M^{-3}_R$ and $I$ is a normalization
factor . The case $N_0>>1$ corresponds to the high  
density regime with many particles in the effective
correlation volume. 

\subsection{Preliminary considerations of the dynamics:}

Before engaging in a full numerical solution of the evolution equations, we
can obtain a clear, qualitative understanding of the main features of the
evolution by looking at the quantum mechanics of the wave functional.

The dynamics is different for the different solutions 
 for $ \Delta_q $ in Eq. (\ref{Deltak}) and can be understood with  simple
 quantum mechanical arguments:   
\subsubsection{Case I}

$\Delta_q=2[n_{q}(0)+\sqrt{n_{q}^2(0)+n_{q}(0)}]$: 

In this case 
\begin{eqnarray}
\Delta_q(0) & \approx & 4 \; n_q(0) >>1 \; \;  \text{for} \; \;
 \frac{|q-q_0|}{\xi} 
 \approx 1 \nonumber \\
  & \approx & 0 \; \;  \text{for}\; \; \frac{|q-q_0|}{\xi}
 >> 1, \label{range}
\end{eqnarray}

and the covariance of the wave functional (\ref{wavefunc0}) is given  by 
\begin{eqnarray}
A_q(0) = \Omega_q(0)  =  \omega_q(1+\Delta_q)& >> & \omega_q \; \; 
\text{for}\; \; \frac{|q-q_0|}{\xi} \approx 1 \nonumber \\
& \approx & \omega_q \; \; 
\text{for}\; \; \frac{|q-q_0|}{\xi} >> 1.
\label{inicova}
\end{eqnarray}

Now, for each wave vector  $ \vec q $ we have a Gaussian wave-function
which is the ground state of a harmonic oscillator with frequency
$\Omega_q(0)$, but 
whose evolution is determined by a Hamiltonian for a harmonic oscillator of
frequency $\omega_q(0)$ at very early times. For the modes $ q $ such that
$\Omega_q >> 
\omega_q(0)$, the wave function is very narrow compared to the
second derivative of the potential and there is a very small probability for 
sampling large amplitude field configurations (i.e. large $ \eta_{\vec
q} $). It is a property of these squeezed states that whereas the 
wave-functional is narrowly localized in field space, it is a wide
distribution in the canonical momentum (conjugate to
the field) basis. 
This wave function will spread out under time 
evolution to obtain a width compatible with 
the frequency $\omega_q$, i.e. the covariance $A_q(\tau)$ will {\em diminish in
time} and the fluctuation 
\begin{equation}
\langle |\vec{\eta}_q|^2 \rangle(\tau) \propto \frac{1}{A_{Rq}(\tau)}
\label{flucu} 
\end{equation}
will {\em increase} in time. This will in turn cause the time dependent
frequency in the Hamiltonian, $\omega_q(\tau)$ to {\em increase} with
time since the frequency and the fluctuations are directly related via
the self-consistency condition. The 
resulting dynamics is then expected to approach an oscillatory regime in which
the width of the wavefunctional and the frequency of the harmonic oscillator
are of the same order. Under these circumstances, there is 
the possibility that for a particular  range of parameters (coupling,
central momentum of  the distribution and particle 
density)  parametric amplification can occur \cite{noneq,boydiss} that could
result in particle production and redistribution of particles as will
be discussed below within an early time analysis. We will see that this case
corresponds to a ``tsunami'' configuration in a theory which is {\em symmetric}
even in the absence of the medium.

\subsubsection{Case II}  
$ \Delta_q=2[n_{q}(0)-\sqrt{n_{q}^2(0)+n_{q}(0)}]$: 

In this case 
\begin{eqnarray}
\Delta_{q}(0)& \approx &  -1+\frac{1}{4n_{q_0}} \;\; \;  \text{for}\; \;
 \frac{|q-q_0|}{\xi} \approx 1  
 \nonumber \\
& \approx & 0 \; \; \text{for}\; \; \frac{|q-q_0|}{\xi} >> 1, \label{delgreat}
\end{eqnarray}
 and the covariance is
\begin{eqnarray}
A_q(0) \approx \frac{\omega_q}{4n_q(0)} & << &  \omega_q \; \; 
\text{for}\; \; \frac{|q-q_0|}{\xi} \approx 1 \nonumber \\
& \approx & \omega_q\; \; \text{for}\; \; \frac{|q-q_0|}{\xi} >> 1. 
\end{eqnarray}
Therefore,  large amplitude field configurations with momenta in the narrow
``tsunami'' shell,
now have a high probability of being realized. As before, the wave function
for each $\vec k$ mode corresponds  
to the ground state of a harmonic oscillator of frequency
$$
\Omega_q(0)={{\omega_q}\over {4n_{q_0}(0)}}
$$ 
which evolves with a Hamiltonian for
a harmonic oscillator with frequency $ \omega_q(0) $. In this case the wave
function  for $ q \approx q_0 $ is spread out over field amplitudes much larger
than
$$
1/\sqrt{\omega_{q_0}}
$$ 
and it is localized in the canonical momentum basis.

Under time evolution the wave function will tend to be squeezed i.e. it
will be forced to 
diminish its width and to become localized inside the potential well. This
implies that the covariance $A_q(\tau)$ will {\em increase} under time
evolution, while the 
fluctuation (\ref{flucu}) and the time dependent frequency $\omega_q(\tau)$ will
{\em decrease}, i.e. the potential `flattens out'.
 
In this case the quantity $g\Sigma(\tau)$ (the renormalized quantum
fluctuations) in the evolution equations (\ref{modeeq}) decreases and
as will be seen below, under certain conditions, can become {\em negative}. 

There is thus a possibility of inducing spinodal 
instabilities in the quantum fluctuations. To see this, consider the case in
which $\varphi \equiv 0$ and the  
effective mass squared ${\cal M}^2_{R\pi}(\tau)=1+g\Sigma(\tau)$ in the equation
(\ref{modeeq}) becomes negative, i.e. when $g\Sigma(\tau) < -1$. The modes for
which $q^2 < |{\cal M}^2_R(\tau)|$  will see an inverted harmonic oscillator
and they will begin to grow almost exponentially, resulting
in copious particle production for these modes as can be seen from the
expression for the particle number as a function of time (\ref{numbexp2}). 

This situation, in which the potential turns into a maximum at the origin
{\em dynamically}, corresponds to symmetry breaking, since the minimum will
be away from the origin.  
In this case the dynamics will result in a re-arrangement of the particle
distribution: spinodal instabilities will arise, long-wavelength
modes will begin to get populated at the expense of the initial non-equilibrium
distribution. The spinodal instabilities will in turn result in an {\em
increase} 
in the fluctuations that will tend to cancel the negative contribution to
$g\Sigma$ from the initial non-equilibrium distribution. Eventually a
stationary regime should ensue in which the instabilities are turned-off and
the distribution 
of particles will be peaked at low momenta.

At this point we want to emphasize that the possibility for the
onset of spinodal
instabilities is purely {\em dynamical}. 
In contrast to previous studies of dynamics in spinodally unstable
situations \cite{wilraj,gavin,boysinglee,boydcc} in which an initially 
symmetric state is evolved with a {\em broken symmetry} Hamiltonian, in
the present case the initial state {\em and the effective Hamiltonian}
are symmetric and the instability is a consequence of the non-equilibrium
dynamics.  
 
 The above analysis of the dynamics, based on the quantum mechanical analogy
 will be shown to be accurate in the next section where we present the details
 of the numerical evolution.

\subsection{Early Time Analysis:}

A more quantitative understanding  of these  cases
can be achieved by studying the early time 
behaviour of the solutions and  setting $\varphi \equiv 0$.

\subsubsection{Case I} 
In this case with $ \Delta_q $ given by (\ref{range})and focusing on
the very early time during which backreaction effects can be ignored,
the  solution to the mode equations (Eq. (\ref{modeeq})) with
the 
initial conditions given by Eq. (\ref{inicon}) is simply a superposition of plane waves with frequency $\omega_q(0)$:

\begin{equation}
\phi^*_q(\tau)\approx \cos(\omega_q(0) \tau)-
i( 1 +  \Delta_q) \sin(\omega_q(0) \tau) \label{formu1}.
\end{equation}
The renormalized quantum fluctuations which are dominated by the modes within
the highly populated momentum shell are given by, 
\begin{equation}
g\Sigma(\tau)
\approx-{g}\int_0^\Lambda q^2 dq\frac{\sin^2(\omega_q(0)\tau)
[1-(1+\Delta_q)^2]}{\omega_q(1+\Delta_q)}\label{shellcontri}.
\end{equation}
If the initial distribution of particles is
sufficiently sharp, a qualitative understanding of the early time
dynamics can be obtained by  a saddle point analysis of the
contribution from the region of large occupation number. 

In this limit $g\Sigma(\tau)$ is approximately given by,

\begin{equation}\label{gSig1}
g\Sigma(\tau) \approx +
\frac{4gN_0}{\omega_{q_0}}\sin^2\left(\omega_{q_0}\tau\right).\label{early1} 
\end{equation} 
The mode equations now become
\begin{equation}
\left[\frac{d^2}{d\tau^2}+q^2+1+\frac{2gN_0}{\omega_{q_0}}-
\frac{2gN_0}{\omega_{q_0}}\cos(2\omega_{q_0}\tau)\right]\phi_q(\tau)=0. 
\label{mathieu}
\end{equation}
This is a Mathieu equation whose solutions are of the Floquet form\cite{abramowitz}. The first and broadest instability band is
centered at the value of $q$ given by 
\begin{equation}
q^2= q^2_0-\frac{2gN_0}{\omega_{q_0}}.\label{reso}
\end{equation}
The width of the unstable band depends on the parameter
\begin{equation}
Q=\frac{gN_0}{\omega^3_{q_0}} \label{unspara}
\end{equation}
and can be read off in reference\cite{abramowitz}. There is a rather small
window of relevant parameters that could allow appreciable parametric amplification. 
Whether the backreaction effects allow the unstable band to remain
under time evolution resulting in large particle production and 
redistribution of particles is a detailed dynamical question that will be studied numerically below.

\subsubsection{Case II} 
The dynamics in this case can be understood by the heuristic
arguments presented below. We work with the solution
$\Delta_q=2[n_q(0)-\sqrt{n_q^2(0)+n_q(0)}]$, which in the limit
$N_0 >>1$ yields (\ref{delgreat})

\begin{eqnarray}
\Delta_q&&\approx -1+\frac{1}{4n_{q_0}}\;\;\;\; \text{for}\;\;\;q \approx
q_0 
\label{peak}\\
&&\approx 0 \;\;\;\; \text{otherwise}.   \label{away}     
\end{eqnarray}
leading now to the following approximate form for the fluctuation at early
times:

\begin{equation}
g\Sigma(\tau) \approx
-\frac{4gN_0}{\omega_{q_0}}{\sin^2(\omega_{q_0}\tau)}\label{early} 
\end{equation} 
when the backreaction effects can be ignored. 
The first feature to note is that $g$ and $N_0$ appear together in such a way
that the effective coupling is now $gN_0$ and hence the physics is
intrinsically non-perturbative when $N_0 \approx 1/g$. This situation
is very similar to that in high temperature field theory wherein the
relevant dimensionless quantity is $T/m(T)$ ($m(T)$ is the temperature
corrected effective mass) and the effective coupling constant for
long-wavelength physics is $\lambda T/m(T)$.  
In this situation the  non-perturbative hard-thermal-loop resummation is
required. 

Secondly the expression for $ g\Sigma(\tau) $ is
always less than or equal to zero. Notice that unlike {\it Case I}, $
g\Sigma(\tau) $ is negative [see Eq.(\ref{gSig1})]. In particular  $
g\Sigma(0) = 0 $  and then  $ g\Sigma(\tau) $ becomes negative i.e. it
begins to {\em decrease}. The fact that the 
fluctuations decrease was exactly what we had expected from the wave functional
analysis presented in the previous section.  

Furthermore we see that when $ 4gN_0/\omega_{q_o}>1 $ at very early times there
will be an unstable band of  wave-vectors. An estimate of the width of the band
can be provided by averaging the time dependence of $g\Sigma$ over one period
of oscillation. This estimate yields the band of wave-vectors 

\begin{equation}
0 < q <  \sqrt{\frac{2gN_0}{\omega_{q_o}}-1}= q_m \label{spinoband}
\end{equation}
 which will become spinodally unstable.
The mode functions for these wavevectors will grow exponentially at
early times and their contribution to the fluctuation $g\Sigma(\tau)$
(\ref{dimless}) will grow -- this is a back-reaction mechanism that will tend
to  shut-off the instabilities.  

This means that if we begin with a
completely $O(N)$ symmetric state i.e. $\varphi(0)=\dot{\varphi}(0)=0$
 and if we choose $N_0$ large
enough such that $1+g\Sigma \approx
1-4 g N_0\sin^2(\omega_{q_0}\tau)/\omega_{q_0} < 0$, spinodal
instabilities will 
be triggered and the symmetry will be spontaneouly broken. 
The condition for spinodal instabilities to appear is given by
\begin{equation}
\frac{4g N_0}{\omega_{q_0}} > 1 \label{spinocond}
\end{equation}
which determines the critical value of the particle number in a correlation
volume in terms of the coupling and the peak momentum of the distribution.

In the preceding sections we provided an intuitive understanding of the
underlying mechanism 
of symmetry breaking in terms of a quantum mechanical analogy.
We now provide an alternative argument to clarify 
the physical mechanism for the dynamical symmetry breaking. The argument
begins with the expression for the `dressed' mass in Eq. (\ref{dressedmass})
which we write in terms of dimensionless quantities as 

\begin{eqnarray}  
M^2_R & = & m^2_R+
g M^2_R \int
\frac{q^2dq}{\omega_q(0)}\left[-\frac{\Delta_q}{1+\Delta_q}\right],
\label{dressedmass2} \\ 
m^2_R & = & m^2_B + g \int \frac{q^2\; dq}{\omega_q(0)} \; . \label{renoma}
\end{eqnarray}
 The second term is dominated by the peak in the initial particle distribution.
Using eqns. (\ref{peak},\ref{away}) and using a saddle point approximation
assuming a sharp distribution, we obtain the relationship 
\begin{equation}
M^2_R\left[1-\frac{4g N_0}{\omega_{q0}}\right] = m^2_R. \label{barebroken}
\end{equation}

Then choosing the effective mass $M^2_R >0$ as we have done
throughout, we see that when the 
condition for spinodal instabilities in Eq. (\ref{spinocond}) is fulfilled
then it must be that $m^2_R <0$. Therefore the renormalized 
mass squared
{\em in the absence of} the medium is negative and the medium effects,
i.e. the  
non-equilibrium distribution of particles dresses this mass making the
effective, medium `dressed'  mass squared positive. Thus in the absence
of medium the potential was a (spontaneous) symmetry breaking potential. The 
initial distribution restores the symmetry at $ \tau=0 $ much in the same
way as in finite temperature field theory at temperatures larger than
the critical temperature. However the initial 
state is strongly out of equilibrium and its time evolution re-distributes the
particles towards low momentum and the spinodal instabilities result from
the squeezing of the quantum state as explained above. 

This situation must be contrasted to that in {\it Case I} above. The same
argument, now applied to {\it Case I} leads to the result

\begin{equation}
M^2_R\left[1+\frac{4g N_0}{\omega_{q0}}\right] = m^2_R. \label{bareunbroken}
\end{equation}

We clearly see that with a positive effective mass, {\it Case I}
corresponds to the 
situation in which the theory was {\em symmetric} even without the medium
effects (i.e. $m^2_R >0$).  

Thus we obtain a physical picture of the different cases: in {\it Case I}
the symmetry was unbroken {\em without} a medium and remains unbroken
when the large density of particles is added. By contrast in {\it Case II}, the
symmetry is spontaneously {\em broken } in 
the absence of a medium, the high density initial state restores the symmetry
in a state out of equilibrium. Under time evolution
the dynamics then redistributes the particles producing
spinodal instabilities and breaking the symmetry. 

We reiterate that the second case represents a novel situation which is in a
sense, contrary to
what happens at high temperature where thermal fluctuations suppress the
possibility of long-wavelength instabilities.

 The issue of symmetry breaking is a subtle one here. If 
we begin with symmetric initial conditions,
$\varphi(0)=\dot{\varphi}(0)=0$, the wavefunctional will always be   
symmetric since the evolution will maintain this symmetry. In order to
test whether the symmetry is spontaneously broken or not, one must
provide an initial state that is slightly asymmetric, with a very small initial
expectation value $\varphi(0)\neq 0$, and follow the subsequent    
time evolution. If the expectation value oscillates around zero,
then the symmetry is {\em not} spontaneously broken since the minimum
of the `dynamical effective potential' is at the origin in field space.
If the expectation value begins rolling away from zero and reaches a
stationary value away from zero then one can
assert that there is a dynamical minimum away from the origin and the
symmetry is spontaneously broken. Thus the test of  symmetry breaking
requires an initial condition with a small value of the order parameter.

\subsection{The Late Time Regime}

 The asymptotic value of the order parameter
 can be obtained by analyzing the full dynamics of the
theory and depends on the initial conditions. This reflects the fact that there
 is no static effective potential description of the physics. However,
 some  information about the asymptotic state (when 
$\dot{\varphi}(\infty)=\ddot{\varphi}(\infty)=0$) can be obtained from the
 equation of  
motion (\ref{zeromodeeq}) by setting $\ddot{\varphi}(\infty)=0$ which yields
 the sum rule \cite{boydiss,erice97}
\begin{equation} 
1+\varphi^2(\infty)+g\Sigma(\infty)=0 \label{sumrule}
\end{equation}   
provided $\varphi(\infty)\neq 0$. 
This sum rule  guarantees that
the pions are the asymptotic massless Goldstone bosons since 
$$
{\cal M}_\pi^2(\tau)= 
m_R^2+\frac{\lambda_R}{2}\left[ \phi^2(\infty)+
\langle\pi^2\rangle_R(\infty)\right]
$$ 
(Eq. (\ref{massoft})) and the sum rule
is a consequence of the Ward identities associated with the global $O(N)$
symmetry.

The non-linear evolution of the mode functions results in a
redistribution of particles within the spinodally unstable band. The
distribution becomes 
more peaked at low momentum and the effective potential flattens 
resulting in a non-perturbatively large distribution  of Goldstone bosons
at low momentum. 

\section{Numerical Analysis}

\subsubsection{\it Case I} We have investigated the possibility of parametric 
amplification in this case in a wide region of parameters but always in the
dense regime $N_0 >>1$ and varying the center of the distribution. We find
numerically that the backreaction effects shut off the parametric instabilities
rather soon allowing only small particle production and redistribution
of particles. Typically the distribution develops peaks but remains
qualitatively unchanged and the dynamics is purely oscillatory. 

\subsubsection{Case II}

 The numerical analysis of the problem involves the solution of the coupled
set of equations (\ref{zeromodeeq}), (\ref{modeeq}) and (\ref{gsigma}) appended
with initial conditions. We choose the zero mode initial conditions to be
$\varphi(0)=10^{-3}, \dot{\varphi}(0)=0$ while the mode functions satisfy
$\phi_q(0)=1;\;\dot{\phi}_q(0)=-i\omega_q(0)(1+\Delta_q)$. Here
$\omega_{q}(0)=\sqrt{q^2+1}$ and
\begin{equation}
\Delta_q=2[n_q(0)-\sqrt{n_q^2(0)+n_q(0)}]\; .
\end{equation}
We have tested the numerics with a momentum cutoff $\Lambda=25$ in units of the
renormalized mass $M_R$ and found that after renormalization 
the numerical results are insensitive to the value of the cutoff 
provided it is chosen to be much larger than the largest wave-vector which
becomes spinodally unstable. The initial particle distribution is chosen to be
\begin{equation}
n_{q}(0)=\frac{N_0}{I}e^{-(q-5)^2}, \label{testdist}
\end{equation} 
where the total initial number of particles is taken to be to be $N_0=2000$,
the coupling is fixed at $g=10^{-2}$ and the initial value of the  
order parameter is taken to be $\varphi(0)=10^{-3}$.

{\bf Results:}

\begin{center}
\begin{figure}[t]
\centerline{\epsfig{file=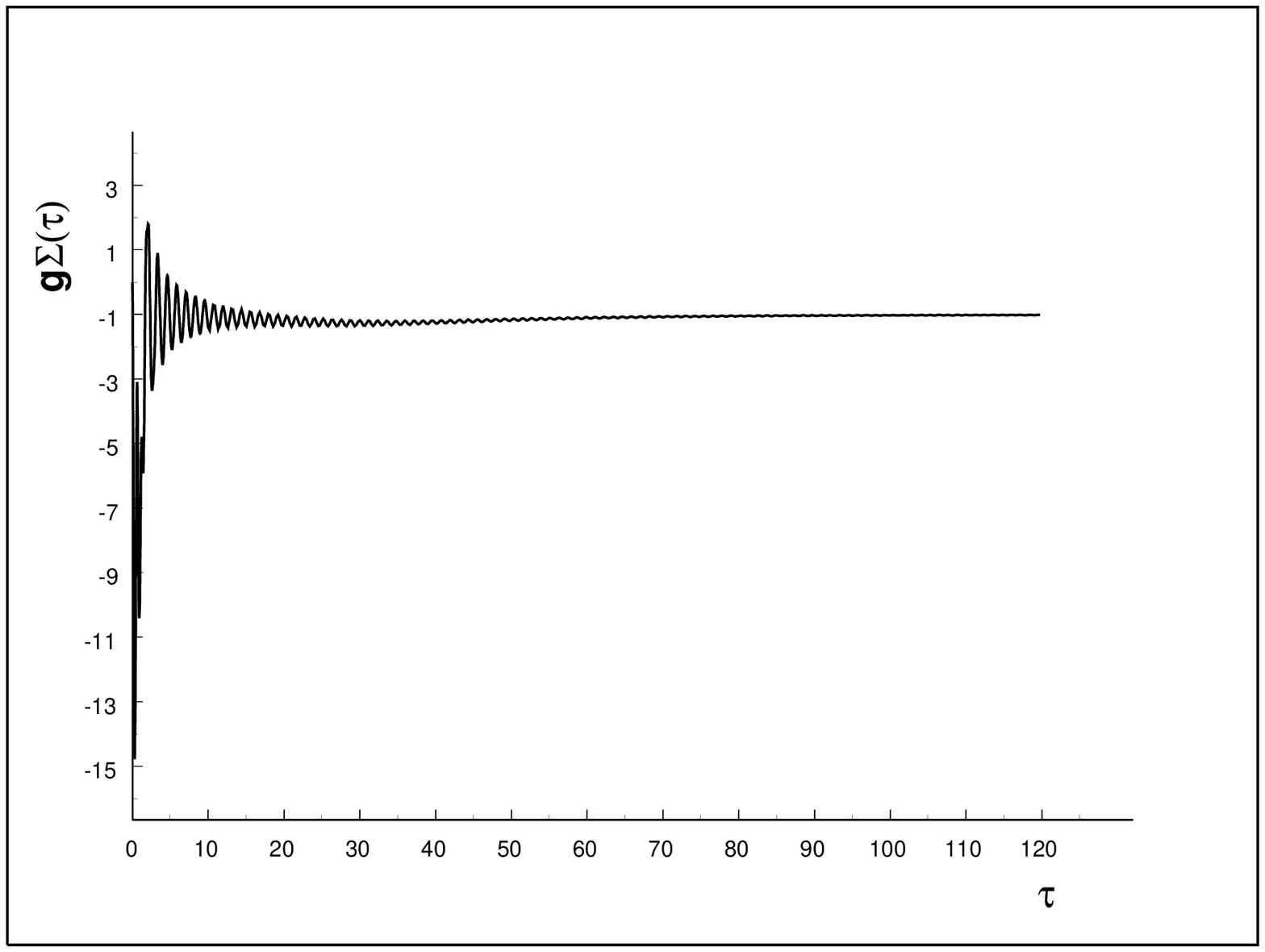,width=3.5in,height=4.5in}
\epsfig{file=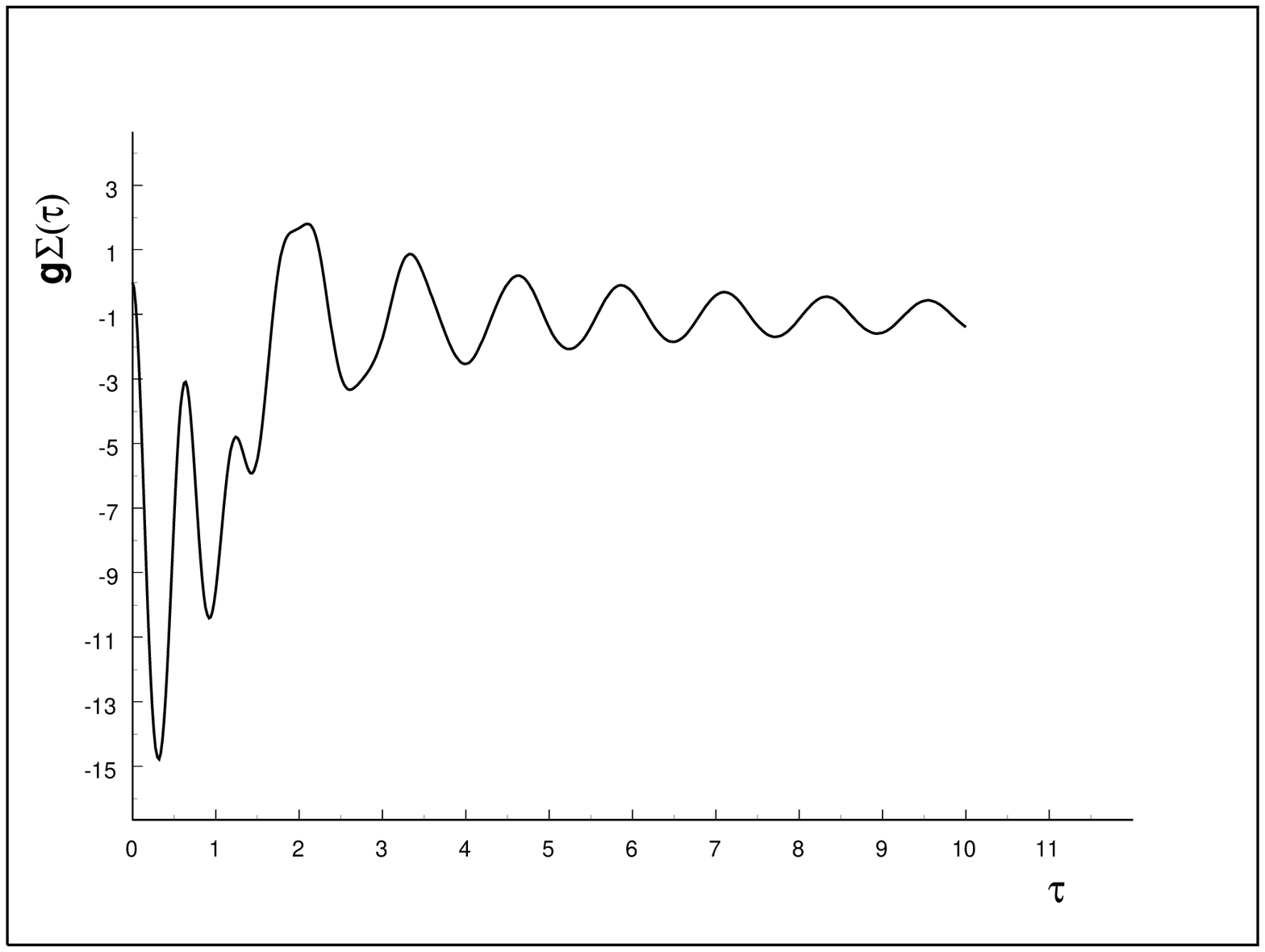,width=3.5in,height=4.5in}}
\centerline{\epsfig{file=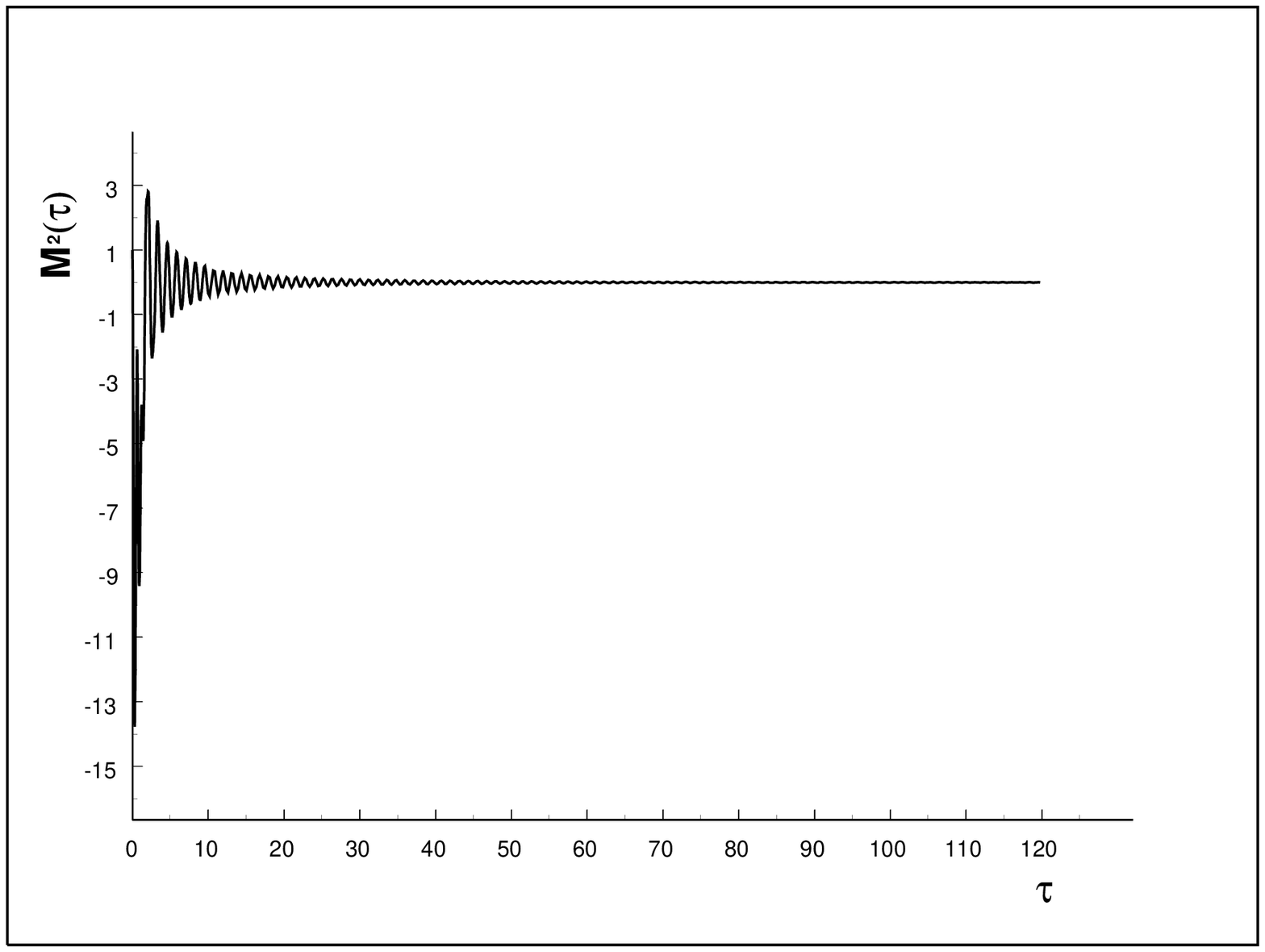,width=3.5in,height=4.5in}}
\caption{$g\Sigma(\tau)$ and ${\cal M}^2(\tau)$ vs. $\tau$ respectively, 
with the initial distribution Eq.(\ref{testdist}), $N_0=2000 \; ; \; 
g=10^{-2}\; ; \; \varphi(0)=10^{-3}\; ; q_0=5$ \label{gsigmass}}
\end{figure}
\end{center}

\begin{center}
\begin{figure}[t]
\centerline{ \epsfig{file=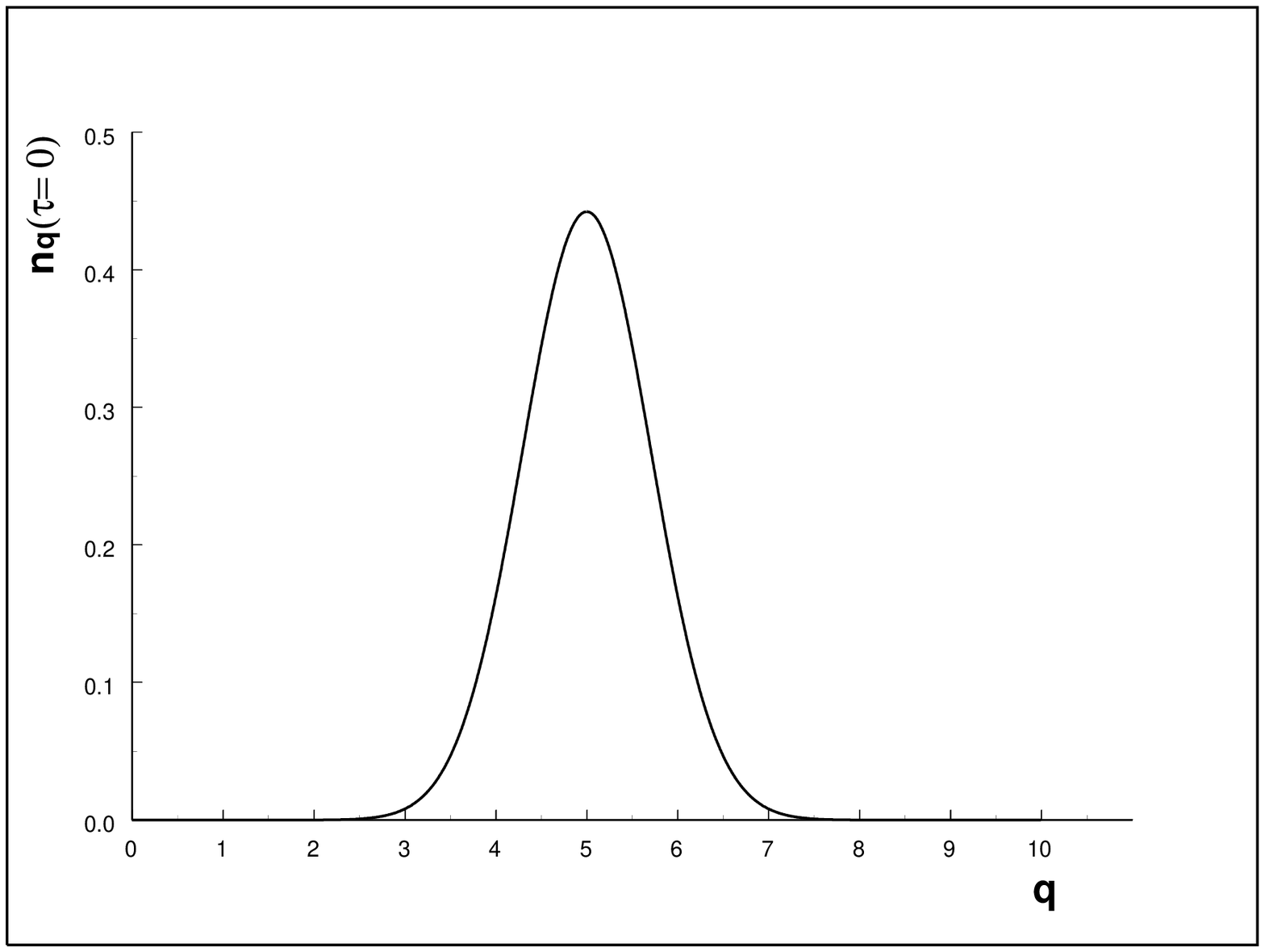,width=3in,height=3in}
\epsfig{file=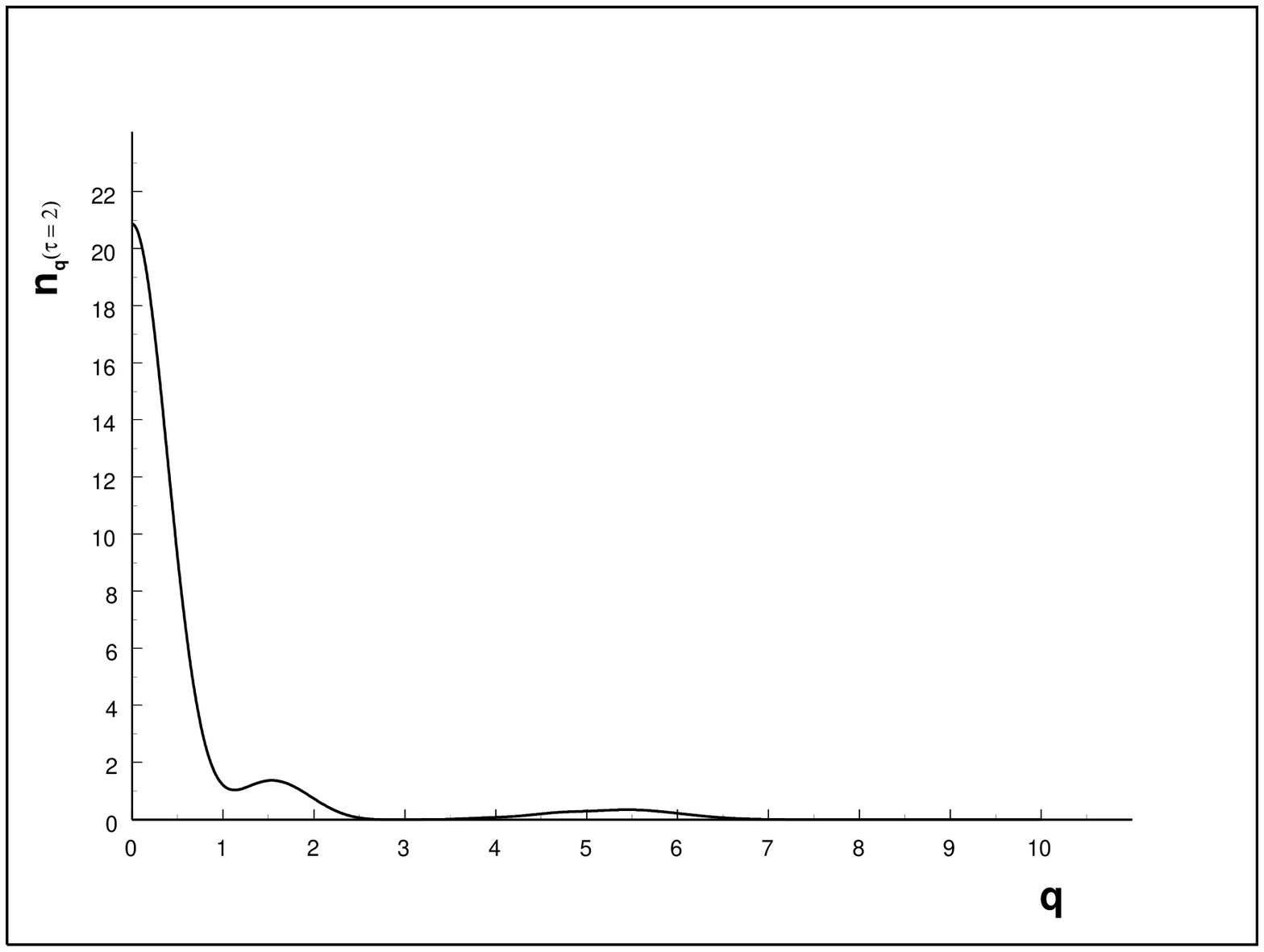,width=3in,height=3in}}
\centerline{\epsfig{file=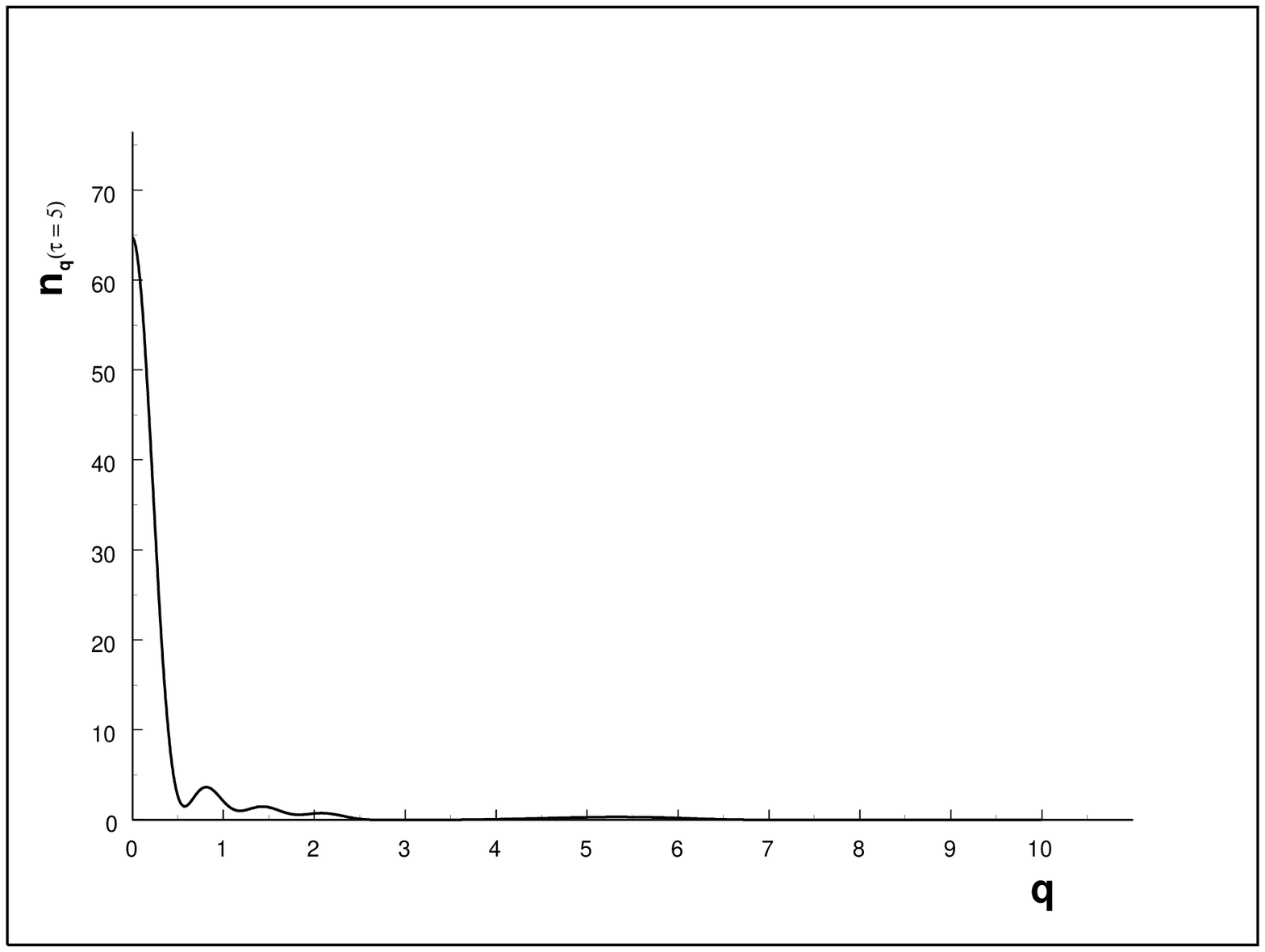,width=3in,height=3in} 
\epsfig{file=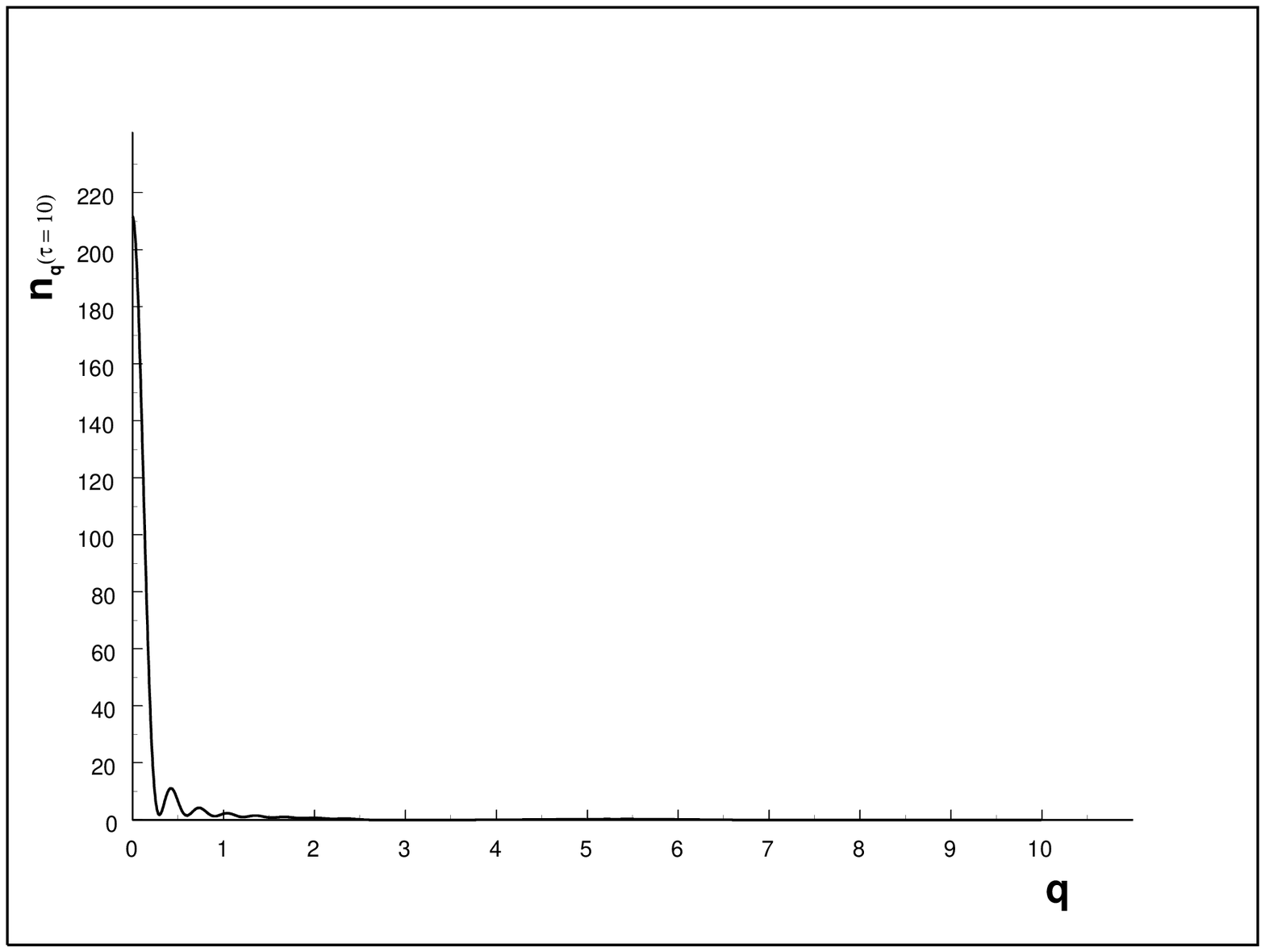,width=3in,height=3in}}
\caption{Distribution function $gn_q$ vs. $q$ for $\tau=0,2,5,10$
respectively with the same parameters as in Fig. (\ref{gsigmass}) \label{nkt}}
\end{figure}
\end{center}

\begin{center}
\begin{figure}[t]
\centerline{\epsfig{file=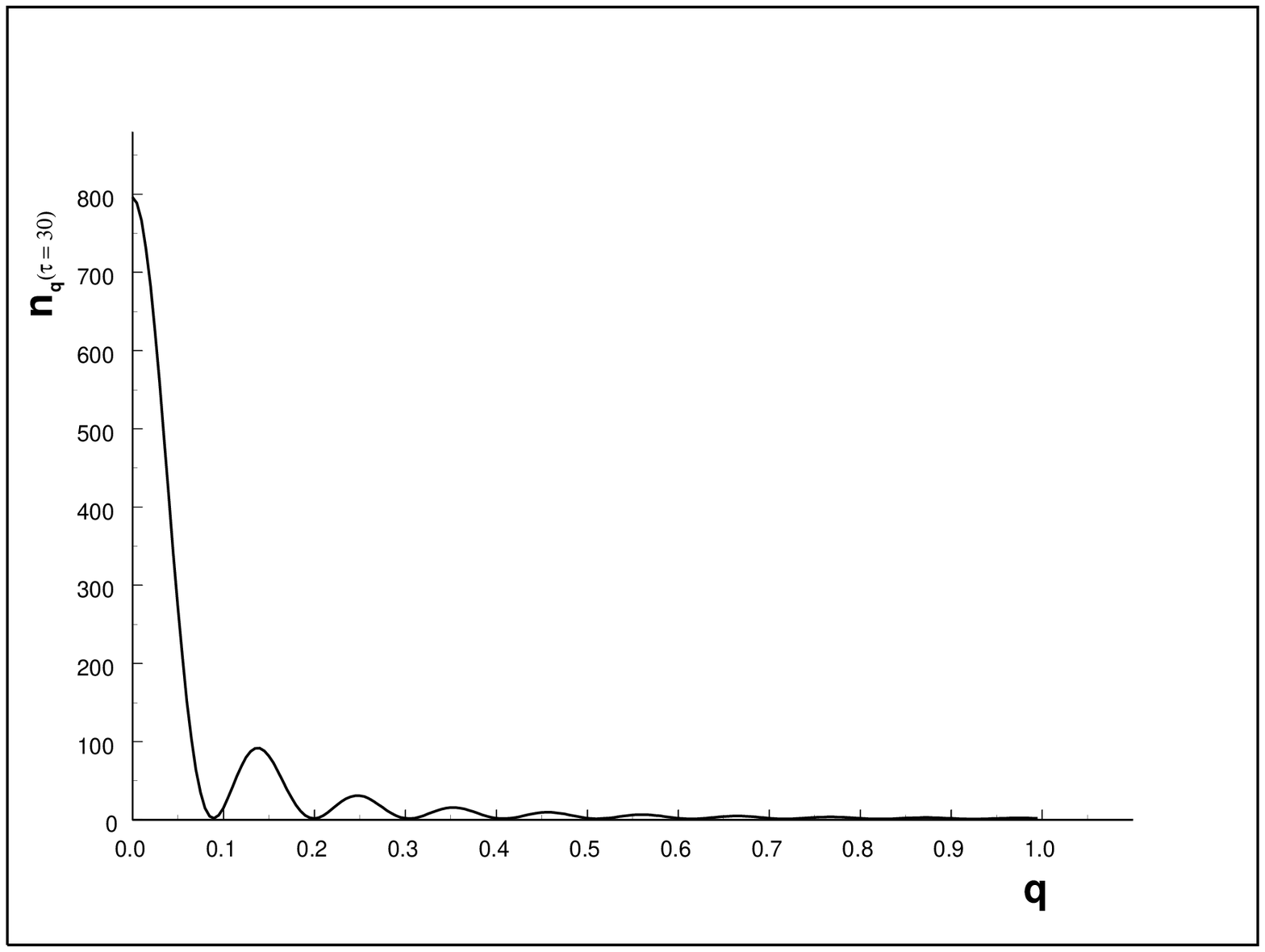,width=3in,height=3in}
 \epsfig{file=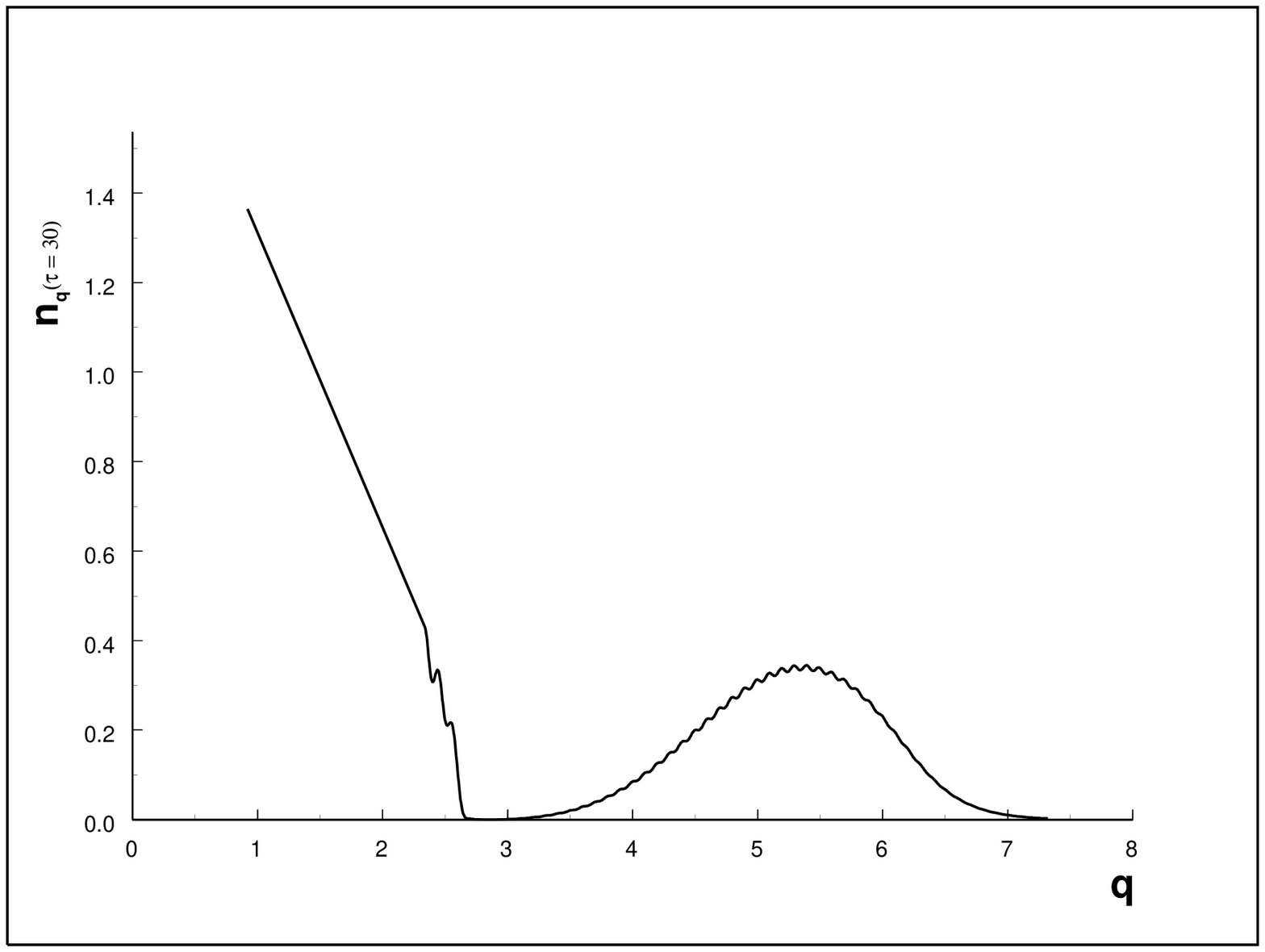,width=3in,height=3in}}
\centerline{\epsfig{file=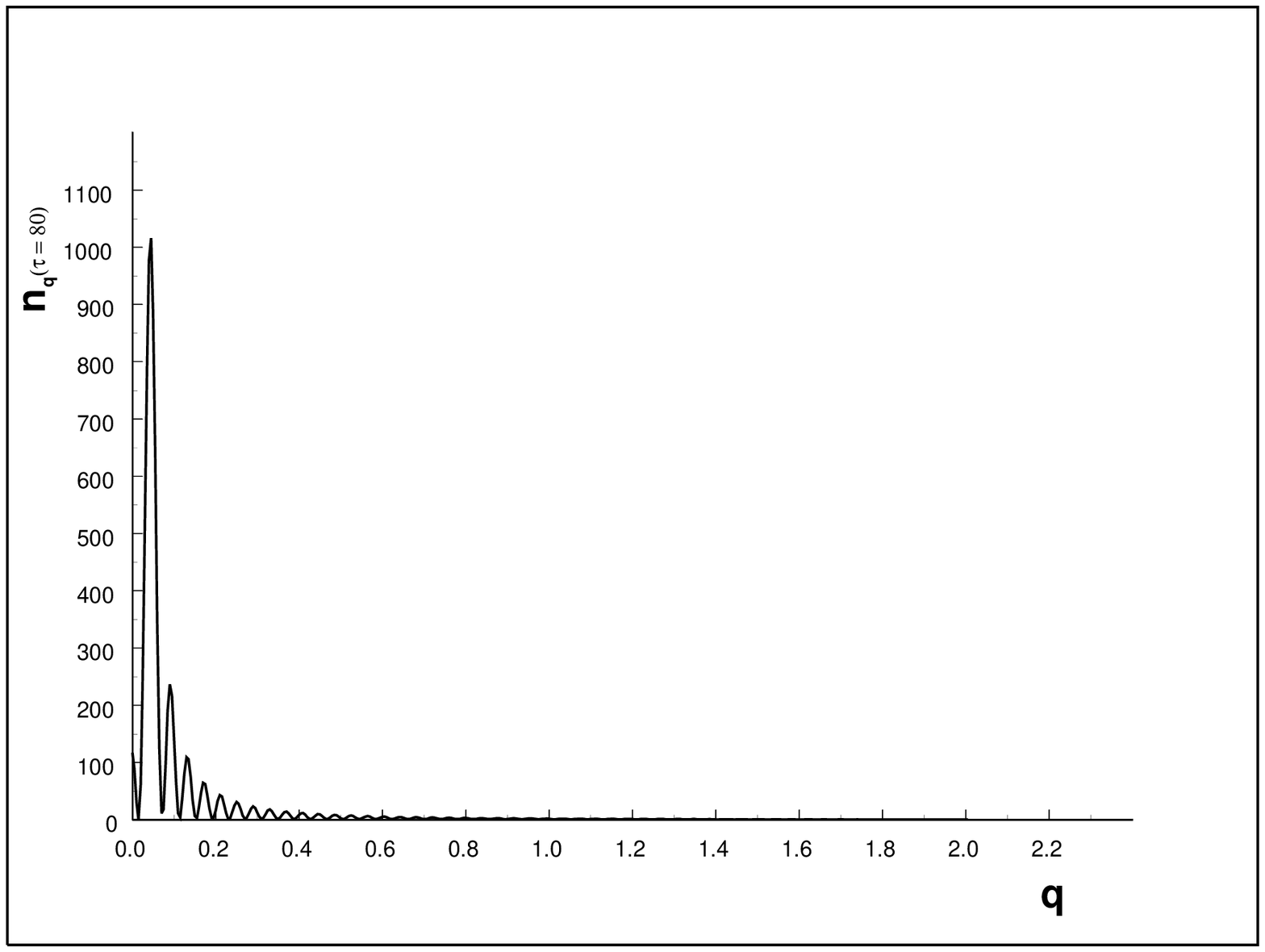,width=3in,height=3in}
\epsfig{file=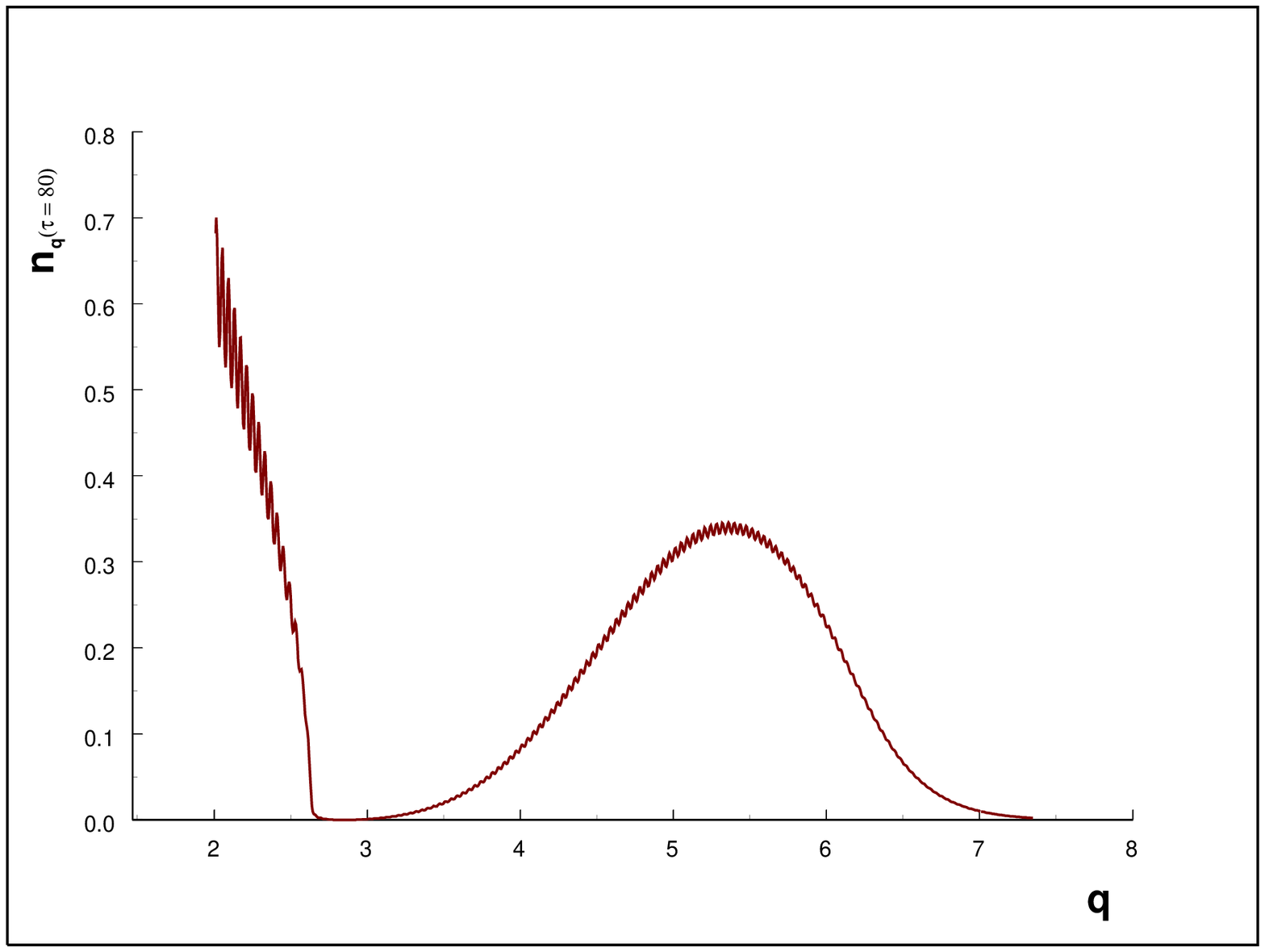,width=3in,height=3in}}
\caption{$gn_q(\tau=30,80)$ vs. $q$.  with the same parameters as in
Fig. (\ref{gsigmass}), zoomed in the region  near $q \approx 0$ and the peak of
the initial distribution  
$q_0=5$.\label{zoomed}}
\end{figure}
\end{center}

\begin{center}
\begin{figure}[t]
 \epsfig{file=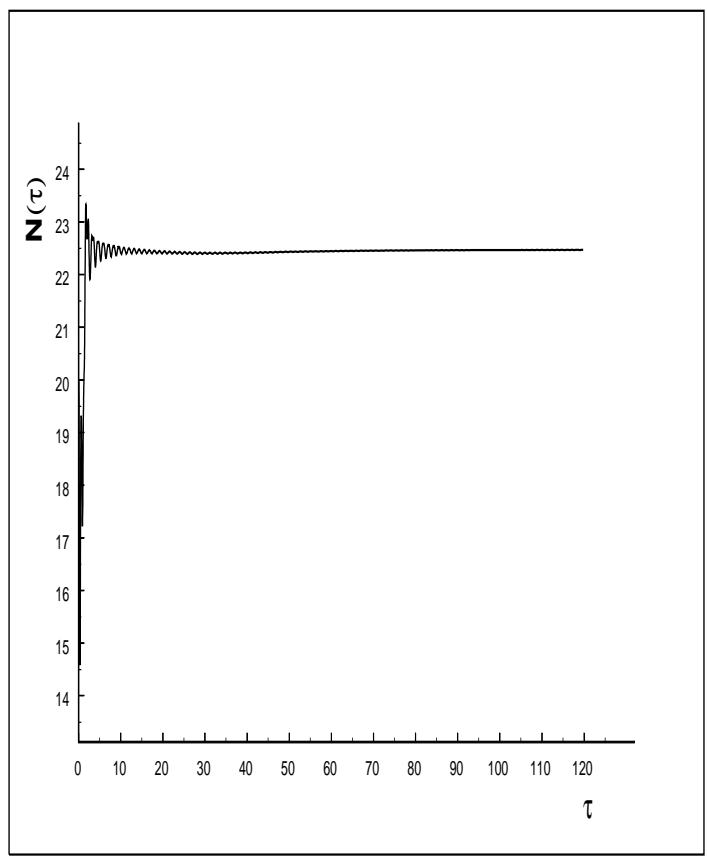,width=4in,height=5in}
\caption{Total number of particles $N(\tau)$ vs. $\tau$.  with the same
parameters as in Fig. (\ref{gsigmass})\label{noft}} 
\end{figure}
\end{center}

Fig.(\ref{gsigmass}) shows $g\Sigma(\tau)$ vs. $\tau$ and the effective
`pion' mass squared ${\cal M}^2_{R\pi}(\tau)$. We see clearly that
spinodal instabilities are produced and the quantitative features of
the dynamics are in agreement with the estimates established for the
early time dynamics given by Eq.(\ref{early}). We see
that the pions become massless, asymptotically. The distribution function
$n_q(\tau)$  
multiplied by the coupling $g$  
is shown in Fig. (\ref{nkt}) at different times, clearly demonstrating 
how the distribution changes in time. As a consequence of the
spinodal instabilities the long-wavelength modes grow exponentially and
the ensuing particle production for these modes populates the band of unstable
modes. 
In particular the amplitudes of the long wavelength modes that become
spinodally unstable grow to be {\em non-perturbatively large} of order
$ 1/g $ and dominate the dynamics completely. 
At earlier times the initial peak in the distribution at $q \approx 5$ 
can still be seen, but at later times it is overwhelmed by the distribution at
long wavelengths. Fig. (\ref{zoomed}) shows a zoom-in 
of the distribution functions ($gn_q$) vs. $q$ at $\tau=30,80$  near $q=0$ and
also 
 near the peak of the initial distribution, around
$q_0 \approx 5$. We see a remnant of the original peak, slighly
shifted to the right but much broader than the initial distribution and
with about half the original amplitude. After $\tau \approx 10$ the
distributions do not vary much in this region of momenta, but they
do vary dramatically at low momenta. Fig. (\ref{noft}) shows the
total number of particles as a function of $\tau$. We see clearly that
initially the total number of particles diminishes because
the fluctuations decrease at early times. The long-wavelength modes
begin to grow because of spinodal instabilities but their contributions are
suppressed 
by phase space. Only when their amplitudes become non-perturbatively large,
is the particle production at long-wavelengths an effective contribution
to the total particle number. When this happens, there is an explosive burst of
particle production following which the total number of particles
remains fairly constant throughout the evolution. After the spinodal
instabilities are shut-off, which for the values chosen for the numerical
evolution correspond to $\tau \approx 2$, the dynamics becomes non-linear.
Whereas during the initial stages the dynamics is in the linear regime,
after backreaction effects have shut-off the spinodal instabilities the
further evolution of the distribution functions is a consequence of the
non-linearities. 

\begin{center}
\begin{figure}[t]
 \epsfig{file=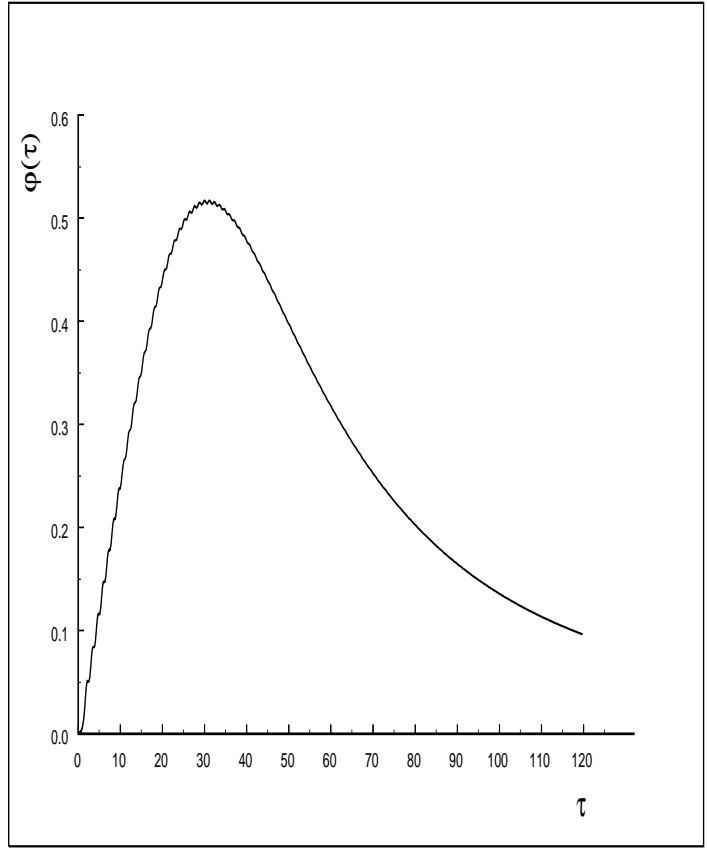,width=4in,height=5in}
\caption{$\varphi(\tau)$ vs. $\tau$.  with the same parameters as in
Fig. (\ref{gsigmass})\label{ordpara}}
\end{figure}
\end{center}

\begin{center}
\begin{figure}[t]
 \epsfig{file=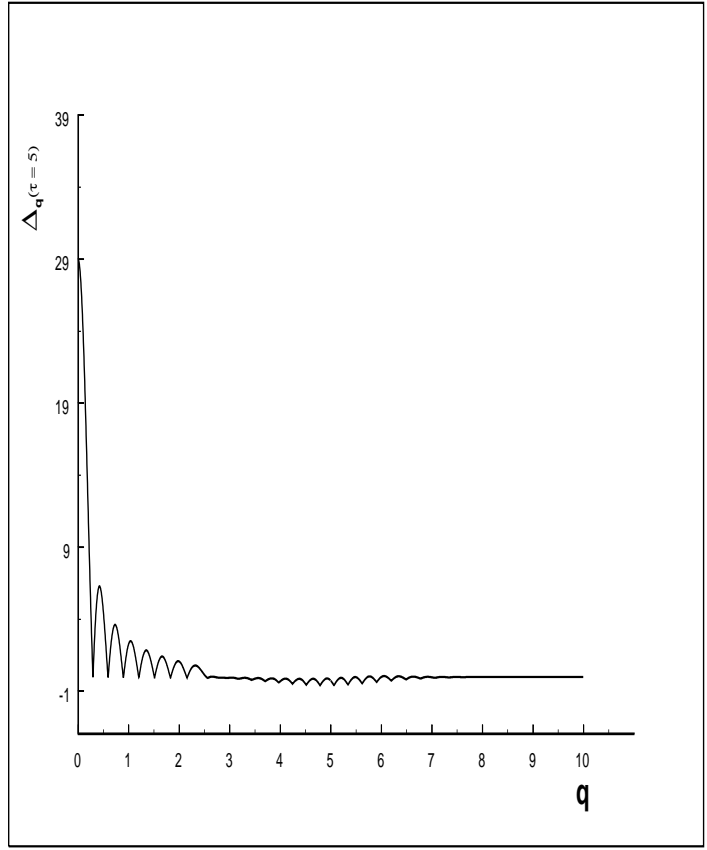,width=4in,height=5in}
\caption{$\Delta_q(\tau=5)=g|\phi_q(\tau=5)|^2-g|\phi_q(\tau=0)|^2$ vs. $q$ for
the same parameters as in Fig. (\ref{gsigmass})\label{modulo}} 
\end{figure}
\end{center}

Fig.(\ref{ordpara}) exhibits one of the clear signals of symmetry breaking. The
order parameter begins very near the 
origin, but once the spinodal instabilities kick in, the origin becomes a
maximum and the order parameter begins to roll away from it. Notice that
the order parameter reaches a very large value, which is the dynamical
turning point of the trajectory, before settling towards a non-zero value.
We find that the value of the turning point and the final value of
the order parameter depend on the initial conditions.  To illustrate 
the non-perturbative growth of modes clearly, we have plotted the quantity 
$g|\phi_q(\tau=5)|^2-g|\phi_q(\tau=0)|^2$ in Fig. (\ref{modulo}) 
which shows how the amplitude of the long wavelength
modes becomes non-perturbatively large and of order $ 1/g $.  

We have also carried out the numerical evolution with $g=10^{-2}, N_0=4000$ and
$g=10^{-3}, N_0=40000$ with the same value of $q_0$ and found the same
quantitative 
behavior, proving that the relevant combination is $gN_0$ as revealed
by the analytic estimates above. We have also confirmed that for $gN_0 <<1$ there 
are no spinodal instabilities and the dynamics is purely oscillatory without
a redistribution of the particles and with no appreciable particle production.
When the peak  of the initial distribution function is {\em beyond} the spinodally
unstable band $q>q_m$ (see Eq.(\ref{spinoband})), 
the original distribution is depleted and broadened somewhat with 
 irregularities and wiggles but remains 
qualitatively unchanged (see fig. \ref{zoomed}). However, when the
peak of the initial distribution is {\em within} the spinodally unstable band  
there is a complete re-distribution of particles towards low momentum.
The original distribution disappears under time evolution and after the
spinodal time only the low momentum modes are populated.

\section{Symmetry Breaking, Energy, Pressure and Equation of State:}

\subsection{ Onset of Bose Condensation:} 
We have seen both from the numerical evolution and from the argument based
on the sum rule (\ref{sumrule}) which is a result of the Ward identities and
Goldstone's theorem, that the effective mass term
vanishes asymptotically. Therefore the asymptotic equation of motion for the
mode 
functions is that of a massless free field. In particular the asymptotic
solution for the $q=0$ mode 
is given by 
\begin{equation}
\phi_0(\tau\rightarrow \infty) = A+B\tau
\end{equation}
where $A$ and $B$ are complex coefficients that can only be obtained from the
full 
time evolution. However because the Wronskian
\begin{equation}
\phi_0(\tau)\dot{\phi}^*_0(\tau)- 
\phi^*_0(\tau)\dot{\phi}_0(\tau)= 2i{W_0} \label{wronski}
\end{equation}
is constant in time, neither $A$ nor $B$ can vanish \cite{thanks}.
This situation must be contrasted with that for the $q\neq 0$ modes whose
asymptotic behavior is of the form
\begin{equation}
\phi_q(\tau \rightarrow \infty) = \alpha_q \; e^{iq\tau}+ \beta_q \;
e^{-iq\tau} \; \;. 
\label{asymqnotzero}
\end{equation}
 This causes
the number of particles at {\em zero momentum} to grow
asymptotically as $\tau^2$ whereas the number saturates for the $q\neq 0$
modes. 
 The three dimensional phase space conspires to
cancel the contribution from the $q=0$ mode to the {\em total}
 number of particles, energy and pressure, which,
from the numerical evolution (see Fig.(4)) are seen to remain constant at 
long times. This situation is very similar to that in 
Bose-Einstein condensation where the excess number of particles at a
fixed temperature goes into the condensate, while the total number of
particles 
outside the condensate is fixed by the temperature and the chemical potential. 
The $q=0$ mode will become macroscopically occupied when $\tau
 \sim\sqrt{V}$
where $V$ is the volume of the system (i.e. the number of particles in
the zero momentum mode becomes of the order of the spatial volume). When this
happens this mode must be isolated and studied separately from the $q \neq 0$ 
modes because its contribution to the momentum integral will be cancelled by
the small phase space at small momentum. Again the situation is very similar 
to the case of the usual Bose-Einstein condensation. Notice that this argument
is independent of a non-vanishing order parameter $\varphi$ and leads to the
identification of the zero momentum mode as a Bose condensate
that signals spontaneous symmetry breaking even 
when the order parameter remains zero. Since the effective mass is
zero we identify the condensing quanta as pions and therefore this mechanism is
a novel form of pion condensation in the absence of direct scattering. 


When scattering is included, beyond the leading order in the large $N$
approximation, 
the formation of the Bose condensate will require a detailed understanding
of the different time scales. The time scale for the  collisionless process
described above must be compared to the time scale for collisional processes
that would tend to deplete the condensate. If spinodal 
instabilities  causing non-perturbative
particle production at low momentum occur on much shorter time scales than
collisional redistribution then we would expect that there will be a 
non-perturbatively large population at low momenta that could be interpreted as
a coherent condensate.  

The spinodal instabilities seen in this article  are similar to those which
lead to the formation of Disoriented Chiral
Condensates\cite{wilraj,gavin,boysinglee,boydcc}. However we emphasize that
unlike most of the previously studied scenarios for DCC formation 
in which a `quench' into the spinodal region was introduced 
{\em ad-hoc}, in the
present situation the spinodal instabilities are of {\em dynamical
origin}. We have studied a situation where the vacuum theory has symmetry
breaking minima   
(with $m_R^2<0$ in Eq. (\ref{barebroken})) but the initial state is highly
excited with the particle density larger
than a critical value leading to a symmetry restored theory in the
medium. However this initial  
state is strongly out of equilibrium and its dynamical evolution automatically
induces
spinodal instabilities.

\subsection{Energy, Pressure and Equation of State:}

As mentioned in the introduction the goal of our study is to understand
the dynamical evolution  of strongly out of equilibrium
states. In the usual investigations of the dynamics of the
quark gluon plasma one uses a hydrodynamic description in which the
energy density, pressure and all the thermodynamic variables depend
only on proper time\cite{bjorken,cooperfry}.
 The hydrodynamic equations are then a consequence
of the conservation laws which are appended with an equation of state to
determine the evolution completely. The hydrodynamic regime corresponds
to the case when the collisional mean free path is shorter than the
wavelength of the hydrodynamic collective modes, and therefore the concept of local thermodynamic
equilibrium is warranted. 

A valid question in the situation that we have
studied in this aricle, 
is whether and when an equation of state is a meaningful
concept. In the leading order in the large $N$ expansion there are no
collisional processes (these arise at ${\cal O}(1/N)$) and therefore
the concept of a hydrodynamic regime in is not
applicable in principle. Furthermore since 
the state considered is spatially homogeneous the pressure will depend
on time rather than on proper time. Since the energy is conserved
and the pressure evolves with time, an equation of state will 
have a meaning only when the evolution has reached the asymptotic regime. 

 The energy density  is given by
\begin{equation}
  \frac{E}{NV}=\frac{1}{2}\dot{\phi}^2+
\frac{1}{2}m^2_B\phi^2 + \frac{\lambda_B}{8} \phi^4  + \frac{1}{4\pi^2}\int
\frac{k^2 dk}{2\Omega_q} 
 \left[|\dot{\phi}_q(\tau)|^2 + \omega^2_q(\tau)
|\phi_q(\tau)|^2\right]-\frac{\lambda_B}{8} \langle \pi^2 \rangle^2.  
\label{enerdens}
\end{equation}

The last term which arises in a consistent large $N$ expansion, is extremely
important in 
that after renormalization 
it provides a {\em negative } contribution which can interpreted as
part of the effective potential\cite{boydiss,noneq}.  
Using the equation of motion for the order parameter and the mode functions it
is straightforward to show that the energy is conserved and  
the last term is necessary to ensure energy conservation.
Since the energy is conserved it can be renormalized by a subtraction at
$\tau=0$, and therefore is finite in terms of the renormalized quantities.  

The pressure is given by the following expression,
\begin{equation}
\frac{p+E}{NV} = \dot{\phi}^2
 + \frac{1}{2\pi^2}\int \frac{k^2 dk}{2\Omega_q}
 \left[|\dot{\phi}_q(\tau)|^2 + \frac{k^2}{3} |\phi_q(\tau)|^2\right].\label{ppluse}
\end{equation}

\begin{center}
\begin{figure}[t]
 \epsfig{file=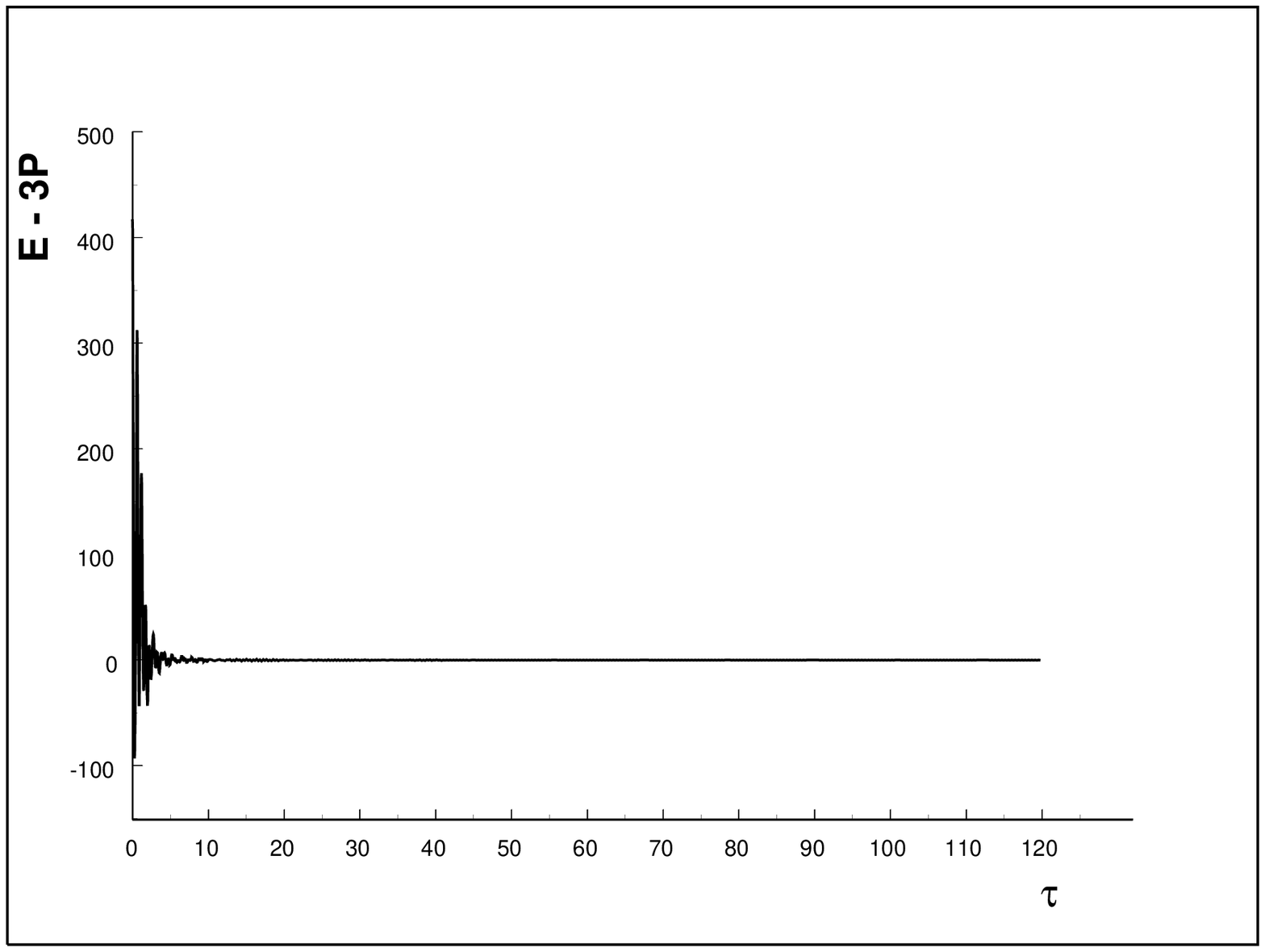,width=4in,height=5in}
\caption{$E-3P$ vs. $\tau$ for the same parameters as in
Fig. (\ref{gsigmass})\label{trace}}  
\end{figure}
\end{center}

Unlike the energy density, the pressure is not a constant of the motion
and needs proper subtractions to render it finite. The detailed expressions for
both the renormalized energy and pressure can be found  
in references \cite{noneq,erice97,baacke}. 
However, rather than computing the total energy
density and pressure, we will study the contributions from the modes
that are highly populated and whose amplitudes become non-perturbatively
large ($\approx 1/g$). Asymptotically, when the effective mass vanishes
and the low momentum modes become highly populated with amplitudes
of $O(1/g)$ the renormalized energy density is given by
(see ref.\cite{noneq,erice97} for the explicit expression of the renormalized
energy density)

\begin{equation}
  \frac{E}{NV}= \frac{1}{4\pi^2}\int^{k_m}_0 \frac{k^2 dk}{2\Omega_q}
 \left[|\dot{\phi}_q(\tau)|^2 + \omega^2_q(\tau) |\phi_q(\tau)|^2\right]
+ {\cal O}(g) \label{enerdensasym}
\end{equation}
where $k_m$ is the largest spinodally unstable wave vector at early times and 
${\cal O}(g)$  
represents terms that are perturbatively small.
Using the asymptotic solutions for the mode functions given by
Eq. (\ref{asymqnotzero}) and neglecting the strongly oscillatory phases 
that average out at long times we obtain
\begin{equation}
  \frac{E}{NV}= \frac{M^4_R}{2\pi^2}\int^{q_m}_0 \frac{q^4 dq}{2W_q}
 \left[|\alpha_q|^2+|\beta_q|^2\right]+{\cal O}(g).
\label{enerdensasym2}
\end{equation}
Similarly, neglecting the contribution of
modes with small amplitudes, we find that the renormalized pressure plus
energy density is given by, 

\begin{equation}
\frac{p+E}{NV} = \frac{4}{3}
  \frac{M^4_R}{2\pi^2}\int^{q_m}_0 \frac{q^2 dq}{2W_q}
 \left[|\alpha_q|^2 +  |\beta_q|^2\right], \label{ppluseasym}
\end{equation}
so that in the asymptotic regime
\begin{equation}
p=\frac{E}{3},
\end{equation}
independent of the particle distribution which is non-thermal. This
is one of the important results of this work. In Fig. (\ref{trace}) we show
the trace of the energy momentum tensor $E-3P$ as a function of time,
for the same value of parameters as for Figs. (1-5). Clearly,
the trace vanishes asymptotically.  

During the early stages of the 
dynamics when spinodal instabilities arise and develop with 
profuse particle production, an equation of state cannot be defined. 
The dynamics cannot be described in terms of hydrodynamic 
evolution. Since the processes under consideration are collisionless, there
is no local thermodynamic equilibrium and an equation of state is ill-defined. 


\section{Conclusions}

We have studied the  evolution of an $O(N)$-symmetric quantum field theory,
prepared in a strongly
out-of-equilibrium initial state. The initial state was characterized by a
particle distribution localized in a thin spherical
shell peaked about a non-zero momentum, a spherical ``tsunami''. 
The formulation of this  scenario resulted from a simplification of the 
idealized `colliding-pancake' description of a heavy-ion
collision. 
For a large density of particles in the initial state, the ensuing dynamics is
non-perturbative and 
consequently we studied the $O(N)$ theory in the leading order in the large $N$
limit which is a systematic non-perturbative approximation scheme . When the
tree-level theory has vacua that spontaneously break the 
symmetry and the 
number of particles within a correlation volume at $t=0$ is so high that
the symmetry is restored initially , spinodal instabilities are then
induced {\em  dynamically} 
resulting in profuse particle production for low momenta.

This situation is to be contrasted with
the usual studies of DCC's where the initial state is assumed to satisfy LTE
(local thermodynamic equilibrium) and the spinodal instabilities are introduced
either via an {\em ad-hoc} quench or via cooling due to hydrodynamic expansion
which is also introduced phenomenologically.

Backreaction of the long wavelength fluctuations eventually shuts off these
instabilities and the nonlinearities redistribute particles  towards
low momenta. We thus find asymptotically in time a novel form of  
pion condensation at low momentum, out of thermal equilibrium.
Furthermore, a macroscopic condensate of the Bose-Einstein type will form 
 at much longer times ($ mt \sim \sqrt{V} $).

When the spinodal instabilities
shut off we find that the asymptotic `quasiparticles' are massless
pions with a non-thermal, non-perturbative distribution function peaked at low
momentum but with an ultrarelativistic equation of state.

We believe that these phenomena point out to very novel 
and non-perturbative mechanisms for particle production and relaxation
that are collisionless, strongly out of local thermodynamic equilibrium and
cannot be described in the early stages via 
a coarse-grained hydrodynamic evolution. These are the result of strongly out
of equilibrium initial states of high density that could potentially 
be of importance in the dynamics of heavy ion collisions at high luminosity
accelerators.   
  
A more realistic treatment, modelling a collision will require an initial
state which breaks the rotational invariance and selects out a beam-axis along
which the colliding pions move in opposite directions. However, the analysis of
such initial conditions is beyond the present numerical capabilities and will
be deferred to a future work. 

An upshot of this study of high density, non-equilibrium particle distributions
is the following tantalizing theoretical question: can one extract a
resummation scheme, or an effective theory akin to the Hard Thermal Loop
effective expansion for arbitrary {\em non-equilibrium, non-thermal}
distributions such as the ``tsunami'' configuration for {\em gauge theories}? 
Possible answers and consequences of such initial states for gauge theories will be 
discussed in a forthcoming article \cite{gaugetsunami}.

\section{Acknowledgements:} 
D. B. thanks the N.S.F for partial support through the grant
awards: PHY-9605186 and  LPTHE for warm hospitality.  R. H.  
 and S. P. K. were supported by DOE grant DE-FG02-91-ER40682. S. P. K. would
like to thank BNL for hospitality during the progress of this work. H. J. de V.
thanks BNL and U. of Pittsburgh for warm hospitality.  
The work of R.D.P. is supported by a DOE grant at
Brookhaven National Laboratory, DE-AC02-76CH00016. 
The authors acknowledge partial support by NATO.

\newpage






\begin{thebibliography}{99}
\bibitem{bjorken} J. D. Bjorken, Phys. Rev. {\bf D 27}, 140 (1983).
\bibitem{book1} L. P. Csernai, ``Introduction to Relativistic Heavy Ion
Collisions'', (John Wiley and Sons, England, 1994).
\bibitem{book2} C. Y. Wong, ``Introduction to High-Energy Heavy Ion
Collisions'', (World Scientific, Singapore, 1994).
\bibitem{muller}J. W. Harris and B. Muller, Annu. Rev. Nucl. Part. Sci.
{\bf 46}, 71 (1996). B. Muller in {\em Particle Production in Highly
Excited Matter}, 
Eds. H.H. Gutbrod and J. Rafelski, NATO ASI series B, vol. 303
(1993). B. Muller, {\em The Physics 
of the Quark Gluon Plasma} Lecture Notes in Physics, Vol. 225 (Springer-Verlag,
Berlin, 
Heidelberg, 1985).
\bibitem{alam} J-e. Alam, S. Raha and B. Sinha, Phys. Rep. {\bf 273}
 , 243 (1996).
\bibitem{meyer} H. Meyer-Ortmanns, Rev. of Mod. Phys. {\bf 68}, 473 (1996).
\bibitem{satz} H. Satz, in {\em Proceedings of the Large Hadron
Collider Workshop} ed. G. Jarlskog and D. Rein (CERN, Geneva), Vol. 1.
page 188; and in {\em Particle Production in Highly Excited Matter},
Eds. H.H. Gutbrod and 
J. Rafelski, NATO ASI series B, vol. 303 (1993). 
\bibitem{wang} X.N. Wang and M. Gyulassy, Phys. Rev. {\bf D44}, 3501 (1991);
Phys. Rev. {\bf D 45}, 844 (1992).
\bibitem{geigmuller} K. Geiger and B. Muller, Nucl. Phys. {\bf B369},
600 (1992). 
\bibitem{geiger} K. Geiger, Phys. Rep. {\bf 258}, 237 (1995);
Phys. Rev. {\bf D46}, 4965  
(1992); Phys. Rev. {\bf D47}, 133 (1993); {\em Quark Gluon Plasma 2}, 
Ed. by R. C. Hwa, (World Scientific, Singapore, 1995).  
\bibitem{eskola} K. J. Eskola and X. N. Wang, Phys. Rev. {\bf D49},
1284 (1994). 
\bibitem{eskola2} K. J. Eskola,hep-ph/9708472 (Aug. 1997).
\bibitem{shuryak} For a recent review see: E. Shuryak in  {\em Quark Gluon
Plasma 2}, Ed. by R. C. Hwa, (World Scientific, Singapore, 1995). 
\bibitem{cooperfry} F. Cooper, G. Frye and E. Schonberg,
Phys. Rev. {\bf D11}, 192 (1975). 
\bibitem{biro} T. S. Biro, H. B. Nielsen and J. Knoll,
Nucl. Phys. {\bf B245}, 449  (1984). 
\bibitem{tube} K. Kajantie and T. Matsui, Phys. Lett. {\bf B164} (1985), 373;
G. Gatoff, A. K. Kerman and T. Matsui, Phys. Rev. {\bf D36}, 114 (114).  
\bibitem{cooper2} Y. Kluger, J. M. Eisenberg, B. Svetitsky, F. Cooper and
E. Mottola, Phys. Rev. Lett. {\bf 67}, 2427 (1991);  Phys. Rev. {\bf D
45}, 4659, (1992); 
Phys. Rev. {\bf D48}, 190 (1993); F. Cooper, in {\em Particle
Production in Highly 
Excited Matter} NATO ASI series B, eds. H. Gutbrod and J. Rafelski,  Vol. 303
(1993).  
\bibitem{anselm1} A. Anselm, Phys. Letters {\bf  B217}, 169 (1989);
A. Anselm and 
M. Ryskin,Phys. Letters {\bf B226}, 482 (1991).
\bibitem{blaizot} J. P. Blaizot and A. Krzywicki, Phys. Rev. {\bf
D46}, 246 (1992); 
Acta Phys.Polon. {\bf 27}, 1687-1702 (1996)
and references therein.
\bibitem{bjorken1} J. D. Bjorken, Int. Jour. of Mod. Phys. {\bf A7},
4189 (1992); Acta Physica Polonica {\bf B23}, 561 (1992). See also
J. D. Bjorken's contribution 
to the proceedings of the ECT workshop on Disoriented Chiral Condensates,
available at http://www.cern.ch/WA98/DCC; G. Amelino Camelia, J. D. Bjorken and
S. E. Larsson, hep-ph/9706530. 
\bibitem{kowalski} K. L. Kowalski and C. C. Taylor, ``Disoriented Chiral
Condensate: A white paper for the Full Acceptance Detector, CWRUTH-92-6.
\bibitem{bjorken2} J. D. Bjorken, K. L. Kowalski and C. C. Taylor: ``Observing
Disoriented Chiral Condensates'', (SLAC-CASE WESTERN preprint 1993)
 hep-ph/9309235 ; ``Baked
Alaska'', (SLAC-PUB-6109) (1993).
\bibitem{revs} For recent reviews on the subject, see: K. Rajagopal, in {\em
Quark Gluon Plasma 2}, 
ed. R. Hwa, World Scientific (1995); S. Gavin, Nucl. Phys.  {\bf A590} (1995),
163c; J. P. Blaizot and A. Krzywicki,hep-ph/9606263 (1996). 
\bibitem{wilraj} K. Rajagopal and F. Wilczek, Nucl. Phys. {\bf B379},
395 (1993), Nucl. Phys. {\bf B404}, 577 (1993). 
\bibitem{boysinglee} D. Boyanovsky, D.-S. Lee and A. Singh,
Phys. Rev. {\bf D48}, 800, (1993).
\bibitem{lattes} C. M. G. Lattes, Y. Fujimoto and S. Hasegawa, Phys. Rep. 
{\bf 65}, 151  (1980); G. J. Alner et al Phys. Rep. {\bf 154}, 247 (1987).
\bibitem{dccexp1} J. D. Bjorken, ``t864 (Minimax): A search fo Disoriented
Chiral Condensate at the Fermilab Collider, hep-ph/9610379 (1996). (See also
the homepage at fnmine.fnal.gov. 
\bibitem{dccexp2} 
See the homepage of the WA98 collaboration at CERN:www.cern.ch/WA98/DCC. 
\bibitem{gavin} S. Gavin, A. Goksch and R. D. Pisarski,
Phys. Rev. Lett., {\bf 72}, 2143 (1994); 
S. Gavin and B. Muller, Phys. Lett. {\bf B329}, 486 (1994).
\bibitem{randrup} J. Randrup, Phys. Rev. Lett. {\bf 77}, (1996), LBL report
38125 (1995); 39328 (1996); hep-ph/9611228 (1996); hep-ph/9612453.
\bibitem{cooperdcc} F. Cooper, Y. Kluger, E. Mottola and J. P. Paz,
Phys. Rev. {\bf D51}, 2377 (1995); Y. Kluger, F. Cooper, E. Mottola, J. P. Paz
and A. Kovner, Nucl. Phys. {\bf A590}, 581c (1995); 
M. A. Lampert, J. F. Dawson and F. Cooper,
Phys. Rev. {\bf D54}, 2213-2221 (1996), F. Cooper, Y. Kluger and E. Mottola,
Phys. Rev. {\bf C 54}, 3298 (1996).
\bibitem{boydcc} D. Boyanovsky, H.J. de Vega and R. Holman,
Phys. Rev. { \bf D51}, 734 (1995). 
\bibitem{rob} Robert D. Pisarski, ``Nonabelian Debye screening,
tsunami waves, and worldline fermions'' 
To appear in the proceedings of the International School of
Astrophysics ``D. Chalonge'', Erice, Italy, Sept. 4-15, 1997; also
based on a talk given at the RIKEN BNL Workshop on ``Non-equilibrium
many body dynamics'', Upton, N.Y., Sept. 23-25, 1997, hep-ph/9710370.
\bibitem{htl} E. Braaten and R. D. Pisarski, Nucl. Phys. {\bf  B337}, 569
(1990);
 E. Braaten and R. D. Pisarski, Nucl. Phys. {\bf B339}, 310
(1990); 
 E. Braaten and R. D. Pisarski, Phys. Rev. Lett. {\bf 64}, 1338
(1990); R. D. Pisarski, Phys. Rev. Lett. {\bf 63}, 1129 (1989); E. Braaten and
R. D. Pisarski, 
Phys. Rev. {\bf D42}, 2156 (1990); J. C. Taylor and S. M. H. Wong,
Nucl. Phys. {\bf B346},
115 (1990); J. Frenkel and J. C. Taylor, {\em ibid} B334, 199, (1990); {\em
ibid}  {\bf B374}, 156 (1992).  
\bibitem{largen1} F. Cooper and E. Mottola, Mod. Phys. Lett. {\bf A2}, 635
(1987).
\bibitem{largen2} F. Cooper, S.-Y. Pi and P. N. Stancioff, Phys. Rev. 
{\bf D34}, 3831 (1986).
\bibitem{largen3} F. Cooper, S. Habib, Y. Kluger, E. Mottola, J. P. Paz,
P. R. Anderson, Phys. Rev. {\bf D50}, 2848 (1994).
\bibitem{frw1} D. Boyanovsky, H.J. de Vega, R. Holman, Phys. Rev. {\bf
D49}, 2769, (1994). 
\bibitem{noneq}D. Boyanovsky, H. J. de Vega, R. Holman and J. F. J. Salgado, 

Phys. Rev. {\bf D54}, 7570 (1996); 

 D. Boyanovsky, D. Cormier, H. J. de Vega and  R. Holman,

Phys. Rev. {\bf D55}, 3373 (1997).
\bibitem{ctp} J. Schwinger, J. Math. Phys. {\bf 2}, 407 (1961);
 K. T. Mahanthappa, Phys. Rev. {\bf 126}, 329 (1962);
 P. M. Bakshi and K. T. Mahanthappa, J. Math. Phys. {\bf 41}, 12
(1963); 
 L. V. Keldysh, JETP {\bf 20}, 1018 (1965);
 K. Chou, Z. Su, B. Hao And L. Yu, Phys. Rep. {\bf 118}, 1
(1985); A. Niemi and G. Semenoff, Ann. of Phys. (NY) {\bf152}, 105 (1984);
N. P. Landsmann and C. G.  van Weert, Phys. Rep. {\bf 145}, 141 (1987);
E. Calzetta and B. L. Hu, Phys. Rev. {\bf D41}, 495 (1990); {\em ibid} 
{\bf D37}, 2838 (1990); J. P. Paz, Phys. Rev. {\bf D41}, 1054 (1990);
 {\em ibid} { D42},  529(1990). 
\bibitem{parwani} R. R. Parwani, Phys. Rev. {\bf D45}, 4695, (1992). 
\bibitem{abramowitz} Handbook of Mathematical Functions, M. Abramowitz and  
I. Stegun, (Dover Publications, N.Y. 1970). 
\bibitem{thanks} We thank E. Mottola and F. Cooper for conversations 
held a long time ago that helped
to understand this argument. 
\bibitem{boydiss} D. Boyanovsky,  H. J. de Vega, R. Holman, D.-S. Lee
and A. Singh, 

Phys. Rev. {\bf D51}, 4419 (1995). 
\bibitem{erice97} D. Boyanovsky, H. J. de Vega and R. Holman, ``Erice Lectures
on Inflationary Cosmology'',  in the Proceedings of the 5th Erice 
Chalonge School on Astrofundamental Physics, Ed. N. Sanchez and
A. Zichichi, (Kluwer) hep-ph-9701304. 
\bibitem{baacke} J. Baacke, K. Heitmann and C. P\"atzold, Phys. Rev. {\bf
D55}, 2320 (1997) and  hep-ph/970627. 
\bibitem{gaugetsunami} D. Boyanovsky, H. J. de Vega, R. Holman, S. P. Kumar and
R. D. Pisarski, in preparation.  


\end{thebibliography}
\end{document}